\documentclass[12pt]{article} 
\usepackage{graphicx,floatflt,amssymb,epsf,rotate}  
\def\gtap{\raisebox{-.4ex}{\rlap{$\sim$}} \raisebox{.4ex}{$>$}}

\begin{document}  
 
\vskip 30pt  
 
\begin{center}  
{\Large \bf Universal Extra-Dimensional models with
boundary localized kinetic terms: Probing at the LHC} \\
\vspace*{1cm}  
\renewcommand{\thefootnote}{\fnsymbol{footnote}}  
{ {\sf Anindya Datta${}^{1}$\footnote{email: adphys@caluniv.ac.in}},  
{\sf Ujjal Kumar Dey${}^{1,2}$\footnote{email: ujjaldey@hri.res.in}},  
{\sf Avirup Shaw${}^{1}$\footnote{email: avirup.cu@gmail.com}}, 
{\sf Amitava Raychaudhuri${}^{1}$\footnote{email: palitprof@gmail.com}}  
} \\  
\vspace{10pt}  
{\small ${}^{1)}$ {\em Department of Physics, University of Calcutta,  
92 Acharya Prafulla Chandra Road, Kolkata 700009, India}}\\ 
  ${}^{2)}$ {\em Harish-Chandra Research Institute, 
Chhatnag Road, Jhunsi, Allahabad  211019, India} \\   
\normalsize

\end{center}

\begin{abstract}  
\noindent

In universal extra-dimensional models a conserved $Z_2$ parity
ensures the stability of the lightest Kaluza-Klein particle, a
potential dark-matter candidate.  Boundary-localized kinetic
terms, in general, do not preserve this symmetry. We examine,  in
the presence of such terms, the single production of Kaluza-Klein
excitations of the neutral electroweak gauge bosons and their
decay to zero-mode fermion-antifermion pairs. We explore how
experiments at the Large Hadron Collider at CERN can constrain
the boundary-localized kinetic terms for different choices of the
compactification radius.

\vskip 5pt \noindent  
\texttt{PACS Nos:~11.10.Kk, 14.80.Rt, 13.85.-t  } \\  
\texttt{Key Words:~~Universal Extra Dimension, Kaluza-Klein, LHC}  
\end{abstract}


\renewcommand{\thesection}{\Roman{section}}  
\setcounter{footnote}{0}  
\renewcommand{\thefootnote}{\arabic{footnote}}  

\section{Introduction}

The Large Hadron Collider (LHC) at CERN has opened a new vista
for exploring particle physics models. Along with the search for
the Higgs boson or the origin of electroweak symmetry breaking,
there is also a keen interest to know about what physics lies
beyond the Standard Model. That there has to be some novelty
around the corner is beyond doubt, important indications being
the issue of naturalness of a relatively light Higgs, the
observed non-zero masses of neutrinos, and the lack of a suitable
candidate for dark matter. What is unknown is the scale at which
this new physics will manifest and the nature of its signal.
There are many models which are being carefully examined:
supersymmetry, extra dimensions, little higgs models, and the
like. It is widely expected that the LHC will soon either find
supporting evidence for one or the other of these models or
tightly constrain them.  In this work we consider a class of
extra-dimensional models where all particles are exposed to an
extra dimension \cite{acd}.  The models under consideration here
can be termed as non-minimal universal extra-dimensions (nmUED)
for reasons which we elaborate in the following. We show here
that  LHC  experiments at 8 TeV may exclude significant
portions of the parameter space of this class of models. 

We explore models with one extra spacelike dimension, $y$, which
is flat and compactified. This coordinate can thus be considered
to run from 0 to $2 \pi R$, where $R$ is the radius of
compactification. All particles -- scalars, spin-1/2 fermions,
and gauge bosons -- are represented by five-dimensional fields
which can be expressed in terms of towers of four-dimensional
Kaluza-Klein (KK) states. The KK states at the $n$-th level
for all particles have the same mass of $n/R$. Further, in
order to incorporate chiral fermions a ${Z}_2$ symmetry ($y
\leftrightarrow -y$) is imposed.  Thus the extra dimension is
compactified on an orbifold $S^1/Z_2$. A translation in $y$ direction 
by an amount $\pi R$ remains a symmetry of this orbifold.  This symmetry leads
to a conserved KK-parity given by $(-1)^n$ where $n$ is the
KK-level. The standard model (SM) particles correspond to $n$ = 0
and are of even parity while the  KK-states of the first level
are odd. The conservation of KK-parity ensures that the lightest
$n = 1$ particle is absolutely stable and hence a potential dark
matter candidate, the Lightest Kaluza-Klein Particle (LKP). This
constitutes what is termed the Universal Extra Dimension (UED)
Model.

The $S^1/Z_2$ compactification leads to two fixed points at $y = 0$
and $y = \pi R$. At these boundary points one can admit additional
interaction terms between the KK-states. In fact, such terms are also
necessitated as counterterms to compensate for loop-induced effects
\cite{georgi} of the 5-dimensional theory. In the minimal Universal
Extra-Dimensional Models (mUED) \cite{cms1, cms2} these terms are
chosen so that the 5-dimensional loop contributions are exactly
compensated at the cutoff scale of the theory $\Lambda$ and the
boundary values of the corrections, e.g., logarithmic corrections to
masses of KK particles, can be taken to be vanishing at the scale
$\Lambda$. Such  contributions can remove the mass degeneracy among
states at the same KK-level $n$.

In this work we allow the boundary terms to be unrestricted by
the special choice in mUED. In this sense the model can be termed
non-minimal UED (nmUED). In mUED the boundary terms are equal at
both fixed points -- a property which may be extended to nmUED.
This will preserve a discrete $Z_2$ symmetry which exchanges $y
\longleftrightarrow (y - \pi R)$.  Here we make a further
departure by allowing the boundary terms to be of different
strengths at the two fixed points or even choosing the boundary
term to be non-zero only at one fixed point. This leads to
breaking of this $Z_2$ symmetry and has far-reaching
consequences\footnote{This is similar to R-Parity violating
interactions in supersymmetry.}. For example, the $n = 1$ excitation of the neutral
gauge bosons, $B^{1}$ and $W_3^1$, can be produced singly at the
LHC and can decay to a zero-mode fermion-antifermion pair. We
examine the impact of different choices of the boundary terms on
the above production and decay rates and explore the prospects of
detecting a signal of KK-particles through this process at the
LHC.

We remark here that there is an alternate way of considering
these theories as defined on a fixed interval in the extra
dimension with boundary conditions applied on the fields
\cite{interval}. The orbifolding alluded to above is disposed off
in this formulation and the boundary conditions offer more
flexibility. Our analysis below can be considered to belong to
either of these formulations.

There have been several explorations of the UED Model and its
variants  with a view to constraining the compactification radius
$R$ and the cut-off $\Lambda$. To start with, an important
feature of these models is that to one-loop order  electroweak
observables have been shown to receive finite
corrections\footnote{This property ceases to hold and sensitivity
of various degrees to the cut-off appears at higher orders.}
\cite{db}. It is therefore meaningful to compare the predictions
of the theory with experimental data and obtain bounds on $R$ and
$\Lambda$. For example, using the muon $(g-2)$ \cite{nath},
flavour changing neutral currents \cite{chk,buras,desh}, $Z \to
b\bar{b}$ decay \cite{santa}, the $\rho$ parameter
\cite{acd,appel-yee}, and other electroweak precision tests
\cite{ewued, precision}, it is found that $R^{-1}~\gtap~300-600$ GeV.  A
relatively low allowed value of $R^{-1}$ encourages the
continuing search for signatures of the model at the Tevatron and
the LHC  \cite{collued} and also in future facilities such as the
ILC or CLIC \cite{ILC}.

In the next section we set up our notations by briefly reviewing
the Universal Extra Dimension model and display the mode
expansions of different fields. This is followed by a discussion
of the nmUED scenario.  Here,  among other things, we present
a new formulation for the fermion fields and point out how
KK-number is violated by boundary-localized kinetic terms.  In
the next section we calculate the $Z_2$-parity violating coupling
involving the first excitation of the neutral electroweak gauge
bosons and a zero-mode fermion-antifermion pair. This coupling
can lead to both the single production of this particle at the
LHC and its decay resulting in a characteristic signal,
relatively background free if the decay product fermion is a
lepton, which is examined next.   Ranges of the
boundary-localized couplings which may be probed by 20
$fb^{-1}$ data from the present 8 TeV LHC run are
discussed. We end with our conclusions.

\section{Universal Extra Dimension}
 
We will be considering a Universal Extra Dimension theory with
one extra dimension. As a typical example, the five-dimensional
Lagrangian for the SM leptons is:
\begin{equation} 
S_{leptons} = \int d^4x ~dy \left\{ \bar{\mathcal{L}}_{i}(x,y) 
(i \Gamma^M D_M )   \mathcal{L}_{i}(x,y) +
\bar{\mathcal{E}}_{i}(x,y) 
(i \Gamma^M D_M )   \mathcal{E}_{i}(x,y)\right\} ,
\end{equation} 
where ${\mathcal{L}_{i}} (\mathcal{E}_{i})$ stands for a
lepton doublet (singlet), $i=1,2,3$ is the generation index,  
and $D_M$ the appropriate covariant derivative. Here the normal
4-dimensional spacetime is denoted by $x (\equiv x^{\mu})$ and
$y \equiv x^4$ is the compactified coordinate. Above, $\Gamma^\mu =
\gamma^\mu$ ($\mu = 0, \ldots, 3)$ and $\Gamma^4 = i \gamma^5$.

In UED the 5-dimensional fields are expressed in terms of the
4-dimensional KK states. Thus in mUED the SM left- and
right-chiral\footnote{The left- and right-chiral projectors are
$(1 - \gamma_5)/2$ and $(1 + \gamma_5)/2$, respectively.} lepton
fields are the zero-modes in the expansions:
\begin{eqnarray} 
\mathcal{L}_{i}(x,y)&=&\frac{\sqrt{2}}{\sqrt{2\pi  
R}}\bigg[{\pmatrix{\nu_i\cr e_i}}_{L}(x)+\sqrt{2}\sum^{\infty}_{n=1}\Big[ 
\mathcal{L}^{(n)}_{iL}(x)\cos\frac{ny}{R}+ 
\mathcal{L}^{(n)}_{iR}(x)\sin\frac{ny}{R}\Big]\bigg], \nonumber \\ 
\mathcal{E}_{i}(x,y)&=&\frac{\sqrt{2}}{\sqrt{2\pi 
R}}\bigg[e_{iR}(x)+\sqrt{2}\sum^{\infty}_{n=1}\Big[ 
\mathcal{E}^{(n)}_{iR}(x)\cos\frac{ny}{R}+ 
\mathcal{E}^{(n)}_{iL}(x)\sin\frac{ny}{R}\Big]\bigg]. 
\label{UEDexpn}
\end{eqnarray} 
These fields satisfy $\mathcal{L}_{i}(x,y) = -\gamma_5
\mathcal{L}_{i}(x,-y)$ and $\mathcal{E}_{i}(x,y) = +\gamma_5 
\mathcal{E}_{i}(x,-y)$ which ensure that the zero-modes are the
SM leptons with the correct chirality. The mass of the $n$-th KK
excitation is $n/R$ irrespective of the other properties of the
field.  Similar expansions for the
five-dimensional quark fields in terms of KK states are quite obvious.

This is the commonly used KK expansion of  five-dimensional
fermion fields in UED. When boundary-localized terms (BLT) come
into play it is convenient to express the four component
5-dimensional fields using two component chiral
spinors\footnote{The Dirac gamma matrices are in the chiral
representation with $\gamma_5 = diag( -I, I)$.}  as
\cite{schwinn}:
\begin{equation} 
\Psi_L(x,y) = \pmatrix{\phi_L(x,y) \cr \chi_L(x,y)} 
=   \sum^{\infty}_{n=0} \pmatrix{\phi_n(x) f_L^n(y) \cr \chi_n(x) g_L^n(y)}
\;\; , 
\label{fiveDL}
\end{equation} 
\begin{equation} 
\Psi_R(x,y) = \pmatrix{\phi_R(x,y) \cr \chi_R(x,y)} 
=   \sum^{\infty}_{n=0}  \pmatrix{\phi_n(x) f_R^n(y) \cr \chi_n(x) g_R^n(y)} 
\;\;  . 
\label{fiveDR}
\end{equation} 
Here we do not indicate the $SU(2)_L$ behaviour of the fields.
If, for example, the above KK expansions are for the electron,
then it has to be borne in mind that $\Psi_L(x,y)$ is one of
the members of the $SU(2)_L$ doublet $\mathcal{L}_{i}(x,y)$ while
$\Psi_R(x,y)$ is the $SU(2)_L$ singlet $\mathcal{E}_{i}(x,y)$ in
eq. (\ref{UEDexpn}). In mUED, eq.  (\ref{UEDexpn}), $f^n_i(y),
~g^n_i(y), ~(i =L,R)$ are either a sine or a cosine function of
$y$. For nmUED this will no longer be the case, as we
discuss later. Further, the mass of the KK-excitations will
deviate from the simple $n/R$ formula and will be solutions of a
transcendental equation.

Needless to say, in UED the scalar and vector boson fields are
also 5-dimensional. The lagrangian for these fields as well as their KK
expansions can be similarly written down.

\section{Boundary-localized terms} 

In nmUED one additionally considers kinetic and mass terms
localized at the fixed points of the orbifold. In this work we
restrict ourselves to boundary-localized kinetic terms (BLKT)
only \cite{Dvali, carena, delAguila, delAguila_STU, flacke, asesh}.

We consider specifically the interaction of quarks and leptons
with electroweak gauge bosons in a 5-dimensional theory with
additional kinetic terms localized at the boundaries at $y = 0$
and $y = \pi R$.  To set the stage, we consider fermion fields
$\Psi_{L,R}$ whose zero-modes are the chiral projections of the
SM fermions.   To our knowledge this treatment of fermion
fields with BLKT given below is new and different from all
earlier ones. In terms of these fields the five-dimensional free
fermion action with boundary-localized kinetic terms is written
as
\cite{schwinn}:
\begin{eqnarray} 
S & = \int d^4x ~dy \left[ \bar{\Psi}_L i \Gamma^M \partial_M \Psi_L 
+ r^a_f \delta(y) {\phi} ^\dagger _L i \bar \sigma^\mu \partial_\mu \phi_L 
+ r^b_f \delta(y - \pi R) {\phi} ^\dagger _L i \bar \sigma^\mu
\partial_\mu \phi_L
\right. \nonumber \\
&  \left. + \bar {\Psi} _R i \Gamma^M \partial_M \Psi_R
+ r^a_f \delta(y) {\chi} ^\dagger _R i {\sigma}^\mu \partial_\mu \chi_R 
+ r^b_f \delta(y - \pi R) {\chi} ^\dagger _R i {\sigma}^\mu
\partial_\mu \chi_R
\right]  ,
\end{eqnarray} 
with $\sigma^\mu \equiv (I, \vec{\sigma})$ and $\bar{\sigma}^\mu
\equiv (I, -\vec{\sigma})$, $\vec{\sigma}$ being the $(2 \times
2)$ Pauli matrices.  Here $r^a_f, r^b_f$ parametrize the strength
of the boundary terms which we choose to be the same for $\Psi_L$
and $\Psi_R$ for illustrative purposes.

Using eq. (\ref{fiveDL}), variation of the above action leads to
coupled equations for the $y$-dependent wave-functions of
$\Psi_L$. These are
\begin{equation}
\left[1 + r^a_f \delta(y) + r^b_f \delta(y - \pi R) \right] m_n f_L^n - 
\partial_y g_L^n = 0,\;\;
m_n g_L^n + \partial_y f_L^n = 0, \; (n = 0,1,2, \ldots).
\end{equation}
Analogously, using eq. (\ref{fiveDR}), for the $y$-dependence of
$\Psi_R$ one obtains
\begin{equation}
\left[1 + r^a_f \delta(y) + r^b_f \delta(y - \pi R) \right] m_n g_R^n + 
\partial_y f_R^n = 0,\;\;
m_n f_R^n - \partial_y g_R^n = 0, \;  (n = 0,1,2, \ldots).
\end{equation}

Eliminating $g_L^n$ and $f_R^n$ one obtains the equations:
\begin{eqnarray}
\partial_y^2 f_L^n &+& \left[1 + r^a_f \delta(y) + r^b_f \delta(y - \pi R) 
\right] m_n^2 f_L^n = 0, \nonumber \\
\partial_y^2 g_R^n &+& \left[1 + r^a_f \delta(y) + r^b_f \delta(y - \pi R)
\right] m_n^2 g_R^n = 0,
\end{eqnarray}
The equations of motion above are of similar form. Below we
discuss the solutions for $f_L$ and $g_L$, denoted by $f$ and $g$
henceforth, which will also apply
{\em mutatis mutandis} for $f_R$ and $g_R$.

The boundary conditions which we impose are \cite{carena}:
\begin{equation}
f^n(y)|_{0^-} = f^n(y)|_{0^+},\;\; f^n(y)|_{\pi R^+} = f^n(y)|_{\pi R^-} , 
\end{equation}
\begin{equation}
\frac{df^n}{dy}\bigg|_{0^+} - \frac{df^n}{dy}\bigg|_{0^-} = -r_f^a
m_n^2 f^n(y)|_{0}, \;\;
\frac{df^n}{dy}\bigg|_{\pi R^+} - \frac{df^n}{dy}\bigg|_{\pi R^-} = -r_f^b
m_n^2 f^n(y)|_{\pi R} .
\end{equation}
We then obtain the solutions:
\begin{eqnarray}
f^n(y) &=& N_n \left[ \cos (m_n y) - \frac{r_f^a m_n}{2} \sin (m_n
y) \right] \;,\;\;  0 \leq y < \pi R,   \nonumber \\ 
f^n(y) &=& N_n \left[ \cos (m_n y) + \frac{r_f^a m_n}{2} \sin (m_n
y) \right] \;,\;\; -\pi R \leq y < 0.
\label{sol1}
\end{eqnarray}
where the masses $m_n$ for  $n = 0,1, \ldots  $ 
satisfy the transcendental equation \cite{carena}:
\begin{equation} 
(r^a_f r^b_f ~m_n^2 - 4) \tan(m_n \pi R)= 2(r^a_f + r^b_f) m_n \;.
\label{trans1}
\end{equation}

The solutions
satisfy the {\em orthonormality} relations:
\begin{equation}
\int dy \left[1 + r^a_f \delta(y) + r^b_f \delta(y - \pi R)
\right] ~f^n(y) ~f^m(y) = \delta^{n m}\;\;.
\end{equation}

The departure of eq. (\ref{sol1}) from eq. (\ref{UEDexpn}) in
that the wavefunctions are  combinations of a sine and a cosine
function rather than any one of them alone and the fact that the
KK masses are solutions of  eq. (\ref{trans1}) rather than simply
$n/R$ are at the root of the novelty of nmUED over the other
versions of the theory.  In our discussions below for fermions
we will take either symmetric boundary-localized terms, i.e.,
$r^a_f = r^b_f
\equiv r_f$ or an extreme asymmetric situation where $r^a_f \neq
0, ~r^b_f = 0$. In the latter limit eq. (\ref{trans1}) becomes
\begin{equation} 
\tan(m_n \pi R)=-\frac{r^a_f m_n}{2} .
\label{trans3f}
\end{equation} 

Let us first pay attention to the solutions $m_n$. We illustrate
a few issues taking as reference the above transcendental
equation, eq. (\ref{trans3f}), which determines $m_n$. Somewhat
similar considerations also apply to eq.  (\ref{trans1}).
Defining $x = m_n \pi R$ and $y = r^a_f/2 \pi R$ the equation
becomes $\tan x = -y \;x$. We require the solutions $x$ as the
{\em real} parameter $y$ varies. It is worth bearing in mind that
irrespective of the value of $y$ one root is $x = 0$ which
corresponds to the zero-mode which receives no corrections from
the BLKT.  Besides the zero-mode, our focus for any $y$ is on
those solutions which correspond to {\em real} $m_n$ and in
particular in this work on the $n = 1$ state which is the smallest
root after the zero-mode. To start with, for vanishing BLKT
parameters (i.e., $y = 0$), the solutions are simply $\pm {n \pi
\over R}$ ($n$ being any integer including 0), which is the
case of basic UED.  With non-zero BLKTs ($y \neq 0$), the $n = 1$ real
root  can be determined graphically for both positive and
negative\footnote{For $y < -1$, i.e., $r^a_f < - 2 \pi R$,  the
squared norm of the solution, $|N_n|^2$, can become negative.
See, for example, eq. (\ref{norm2}) below for $n = 0 \Rightarrow
m_n = 0$.  In the following we will consider $y > -1$.} $y$.
They correspond one-to-one with the solutions in the absence of
BLKT but are always smaller (larger) than $n/R$ for the $n$th
mode for positive (negative) $y$.  It is noteworthy that in the
mUED model, where the radiative corrections are calculated in the
framework of the Standard Model gauge group, the mass of the
$n$th mode is more than  $n/R$ (i.e., as for $y < 0$) and is in a
one-to-one correspondence with the mass spectra of KK-modes
\cite{cms1}. Here we focus on the real roots of eqs.
(\ref{trans1}) and (\ref{trans3f}) for positive as well as
negative values of $y$.

Over and above the real roots of eq. (\ref{trans3f}),
for $0 > y > -1$ one also has complex roots in conjugate
pairs\footnote{For $n = 0$ these complex roots are pure imginary
(for $0> r^a_f > - 2\pi R \Rightarrow 0 > y > -1$); a point noted
earlier \cite{delAguila}.}.  These complex solutions for $m_n$
are unphysical. The inclusion of a BLKT, however small, should
not enhance the multiplicity of the Kaluza-Klein states -- the
real solutions having already matched the count.

The constant $N_n$ can be determined from orthonormality  
and is 
\begin{equation}
 N_n = \sqrt{\frac{2}{\pi R}}\left[ \frac{1}{\sqrt{1 + \frac{r_f^2 m_n^2}{4} 
+ \frac{r_f}{\pi R}}}\right],
\label{norm1}
\end{equation}
for symmetric boundary terms.

For the asymmetric case when $r^b_f = 0$ and we use $r^a_f \equiv r_f$ one has 
\begin{equation}
 N_n = \sqrt{\frac{2}{\pi R}}\left[ \frac{1}{\sqrt{1 + \frac{r_f^2 m_n^2}{4} 
+ \frac{r_f}{2 \pi R}}}\right].
\label{norm2}
\end{equation}

In our work  we will deal only with the zero-modes and the $n =
1$ excitations of the five-dimensional fermion fields.
 
Now we turn to the five-dimensional gauge field, $G_N ~(N = 0
\ldots 4)$, where $G$
can be either $W_3$ or $B$. The  action
with boundary kinetic terms can be similarly written as
\begin{equation} 
S = -\frac{1}{4}\int d^4 x \;dy \left[ F_{MN} F^{MN}
+ r_G^a \delta(y) F_{\mu\nu} F^{\mu \nu}
+ r_G^b \delta(y - \pi R) F_{\mu\nu} F^{\mu \nu} \right]  ,
\end{equation} 
where $F_{MN} = (\partial_M G_N - \partial_N G_M)$  and $r_G^a,
~r_G^b$, the strengths of the boundary terms which are varied in our
analysis\footnote{In the remainder of this section we drop the
subscript $G$ on the brane-localized couplings for the gauge
boson: $r^a \equiv r_G^a$ and $r^b \equiv r_G^b$.}.  We will
also comment on an extreme asymmetric case, $r^a \neq 0$ and $r^b = 0$.

The expansion of the gauge  field will be: 
\begin{equation} 
G_{\mu}(x,y)=\sum^{\infty}_{n=0}G_{\mu}^{(n)}(x) a^n(y),~~~~ 
G_4(x,y) = \sum^{\infty}_{n=0}G_4^{(n)}(x) b^n(y),  
\end{equation} 
where the functions $a^n(y)$ and $b^n(y)$ are determined by the
boundary conditions as discussed below. It is convenient to make
the gauge choice: $G_4 = 0$.

The functions $a^n(y)$ satisfy:  
\begin{equation} 
\partial_y^2 a^n(y) + \left[1 + r^a \delta(y) + r^b \delta(y - \pi R)
\right]m_n^2 a^n(y) = 0 .
\label{sol2}
\end{equation} 
We use
the boundary conditions
\begin{equation}
a^n(y)|_{0^-} = a^n(y)|_{0^+},\;\; a^n(y)|_{\pi R^+} = a^n(y)|_{\pi R^-} , 
\end{equation}
\begin{equation}
\frac{da^n}{dy}\bigg|_{0^+} - \frac{da^n}{dy}\bigg|_{0^-} = 
-r^a m_n^2 a^n(y)|_{0}, \;\;
\frac{da^n}{dy}\bigg|_{\pi R^+} - \frac{da^n}{dy}\bigg|_{\pi R^-} = -r^b
m_n^2 a^n(y)|_{\pi R} .
\end{equation}
which imply that the masses $m_n$ are solutions of:
\begin{equation}
(r^{a} r^{b} m_{n}^{2}-4)~\tan \left(m_{n}\pi R\right) = 2(r^{a}+ r^b)
m_{n} \; .
\label{trans2}
\end{equation}

\begin{figure}[thb] 
{\hskip 1.5cm}
\includegraphics[width=4.8cm,height=5.0cm,angle=270]{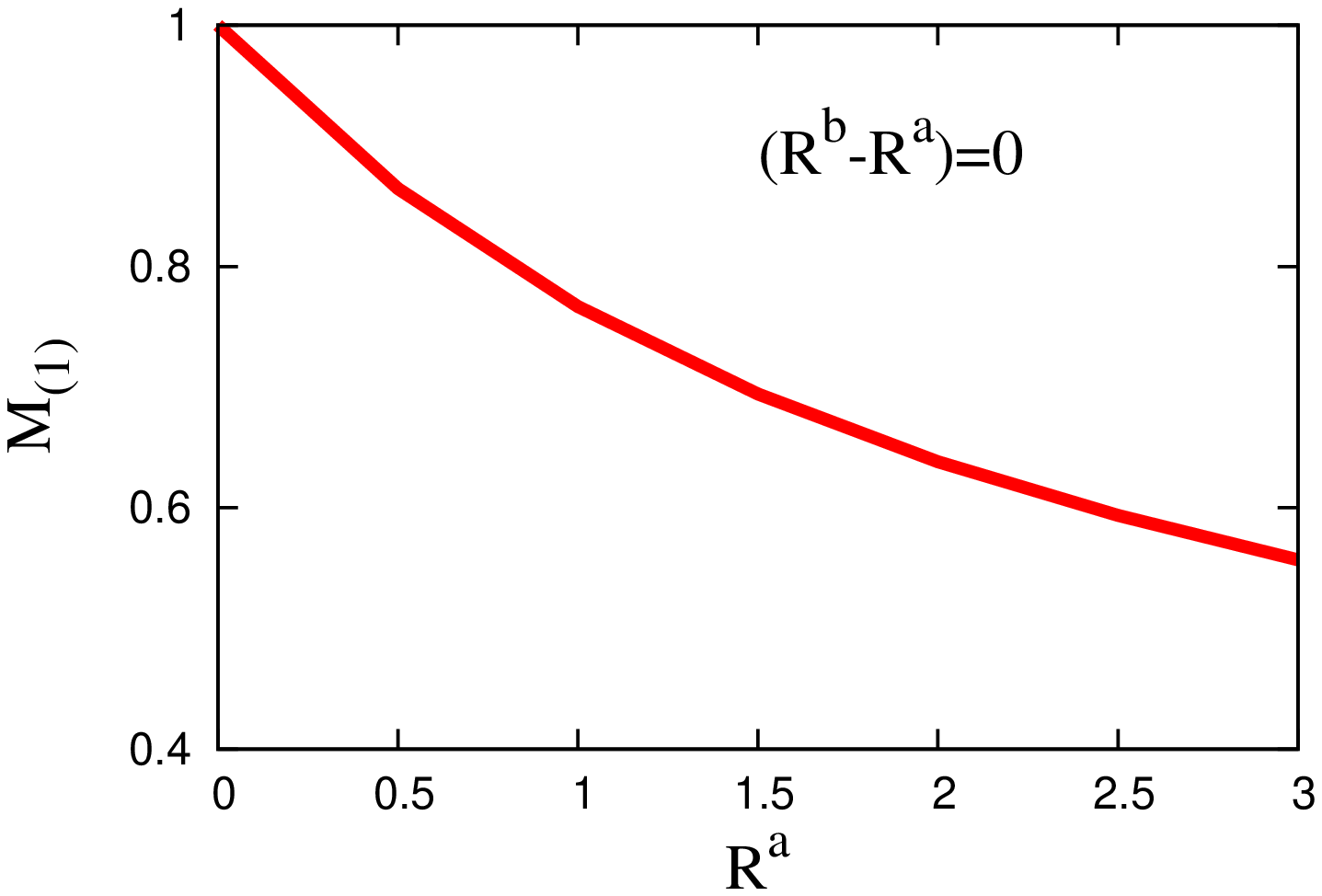} 
{\vskip -4.8cm}
{\hskip 6.8cm}
\includegraphics[width=4.8cm,height=4.8cm, angle=0]{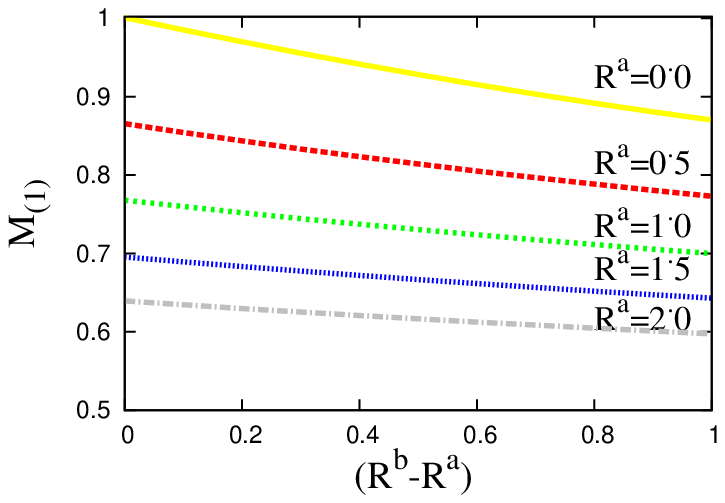}
\caption{Variation of $M_{(1)} \equiv m^{(1)} R$ with the BLKT
strength $R^a \equiv r^a/R$ when $R^a = R^b$ (left) and variation
of $M_{(1)}$  with ${\Delta r \over R} \equiv (R^b - R^a)$ for
different values of $R^a$ (right).  The left panel applies to
fermions, gauge bosons, and Higgs scalars when the BLKTs are
symmetric. Note that larger $R^a$ yields a smaller mass
in both panels. `$a$' and `$b$' correspond to the fixed points
$y = 0$ and $\pi R$, respectively.} 
\label{KKmass} 
\end{figure} 

The roots of the above transcendental equation, which may be
obtained numerically, are the extra-dimensional contributions,
$m_{G}^{(n)} \equiv m_n$, to the masses of the KK
modes\footnote{As discussed below, the KK modes also receive a
contribution to their masses from spontaneous breaking of the
electroweak symmetry.} of the $B$ and $W$.  In our discussion
below we find it convenient to use $\Delta r=r^{b}-r^{a}$.  

Eq. (\ref{trans1}) satisfied by KK fermions and eq.
(\ref{trans2}) valid for KK gauge bosons are of identical form.
It applies to the Higgs scalars as well.  We discuss the
solutions together bearing in mind that for fermions and the
Higgs scalars we take the BLKTs at the two fixed points to be always
equal, i.e., $\Delta r = 0$. In this work we are interested only
in the $n=1$ KK modes.  In fig. \ref{KKmass}, we plot the
dimensionless quantity $M_{(1)} \equiv m^{(1)} R$. In the left
panel we show the variation of $M_{(1)}$ with $R^a \equiv r^a/R$
in the symmetric limit when $\Delta r = 0$ and applies to the
$n$ = 1 fermion and Higgs boson states. It also gives the
extra-dimensional contribution to the  gauge boson mass,
$m_{G}^{(1)}$, in the special case when the BLKTs are symmetric.
When $R^a = 0$, i.e., no BLKT at all, one gets $m^{(1)} =
R^{-1}$, as expected.  Keeping $R^a = R^b$,  one finds that
$m^{(1)}$ monotonically decreases as the BLKT strength $R^a$
increases.  In the right panel, $M_{(1)}$
is displayed as a function of the asymmetry parameter $(R^b -
R^a)$, for several choices of $R^a$.  When $R^a \neq R^b$,
for a fixed $R^a$, $m^{(1)}$ falls as $\Delta r$ grows but this
reduction is not very steep for the range of $\Delta r$ that we
consider. In the right panel the range of $(R^b - R^a)$ and the
values of $R^a$ are chosen to match with those which have been
used later. Notice that the mass of the $n$ = 1 state for a
particular $R^a$ always remains more than that corresponding to
any larger $R^a$ for the entire variation of $(R^b - R^a)$. So,
notwithstanding the value of $(R^b - R^a)$, the ordering of
masses within the $n$ = 1 level can be determined just on the
basis of the corresponding $R^a$s.  This will be useful later in
deciding which state is the LKP.

In the extreme asymmetric case that we consider ($r^a \neq 0, \;
r^b = 0$) for the fermions, gauge bosons, and the Higgs scalars
while eq.  (\ref{sol2}) continues to hold, (\ref{trans2}) reduces
to
\begin{equation} 
\tan(m_n \pi R)=-\frac{r^a m_n}{2} .
\label{trans3}
\end{equation}

\begin{figure}[thb] 
\begin{center} 
\includegraphics[width=4.3cm,height=4.8cm,angle=270]{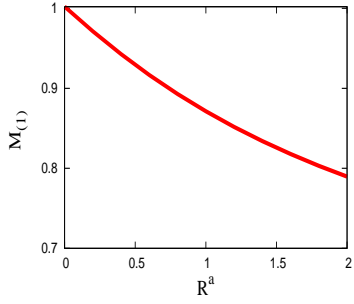} 
\caption{ Variation of $M_{(1)}$ with  $R^a$ when the BLKT is
present only at the $y = 0$ fixed point, i.e., $R^b = 0$. The
results are valid for fermions, gauge bosons, and Higgs scalars.}
\label{KKmassS} 
\end{center} 
\end{figure}

In fig. \ref{KKmassS}, we show the variation of $M_{(1)}$ with
$R^a$  when the BLKT is present only at $y = 0$. Here
again $m^{(1)}$ equals $R^{-1}$ when $R^a = 0$ and
falls with increasing $R^a$ asymptotically approaching the
value $0.5 R^{-1}$.

It is to be noted that the mass $M_{(1)}$ is determined entirely
by the BLKT parametrs, $R^a, R^b$ and the compactification radius
$R$. The gauge coupling is not involved. Therefore this
discussion applies for both $W_3^1$ and $B^1$ so long as the
appropriate BLKT parameters are used.

Before proceeding further it may be worthwhile to identify the
neutral electroweak gauge boson mass eigenstates.  The
electroweak gauge boson eigenstates in a five-dimensional theory
with BLKT have been discussed in  \cite{flacke}. We have checked
that for the range of BLKT parameters which we entertain the
mixing between states of different KK level, $n$, is very small
and may be ignored. Further, only if the BLKT parameters for the $B$
and $W$ gauge bosons are equal or very nearly equal the mixing
between $B^1$ and $W_3^1$ is substantial, it being equal to the
zero-mode weak mixing angle in the case of equality. If $(r_B -
r_W)/R$ is as small as 0.1 this mixing is already negligible.
This can be checked from the mass matrix which we now discuss.

The mass matrix for the $n=1$ neutral electroweak gauge bosons
receives contributions from two sources: one originates from the
spontaneous breaking of the electroweak symmetry and the
other due to the  extra-dimensional contribution discussed above.
When taken together one has the mass matrix:
\begin{equation}
M_{W_{3}^{1}B^{1}} = 
\begin{pmatrix}{
\frac{g^2v^{2}}{4} \frac{S_{W}}{S_H} 
I_{W_{3}W_{3}}+{{m^{(1)}_{W_{3}}}^2}
&
-\frac{gg^{\prime}v^{2}}{4} \frac{\sqrt{S_{W}S_{B}}}{S_H}
I_{W_{3}B}\cr
& \cr
-\frac{gg^{\prime}v^{2}}{4} \frac{\sqrt{S_{W}S_{B}}}{S_H}
I_{W_{3}B}&
\frac{g^{\prime 2}v^{2}}{4} \frac{S_{B}}{S_H}
I_{BB}+{{m^{(1)}_B}^2}
}\end{pmatrix}, \nonumber
\label{mmatrix}
\end{equation}
where
\begin{equation} 
I_{ij}=\int^{\pi R}_{0} 
\left(1+r_{h}\{\delta(y)+\delta(y-\pi R)\}\right)
a^1_{i}(y) a^1_{j}(y)  dy, \;\; (i, j = W_{3},B)  .
\end{equation}

Here $r_{h}$ is the strength of the BLKT for the  Higgs
scalar, taken to be equal\footnote{When we consider the
asymmetric option of BLKTs at one fixed point only, 
the same is the case for the Higgs scalar as well.} at $y=0$ and $\pi R$.
As usual, $R_h = r_h/R$. Also,
\begin{equation} 
a^1_{W_{3}}(y) =
N_{W_{3}}^{(1)}\left[\cos\left({m^{(1)}_{W_{3}}}y\right) -
\frac{r^a_{W}y}{2}\sin\left({m^{(1)}_{W_{3}}}y\right)\right], \nonumber
\end{equation}
and
\begin{equation} 
a^1_{B} (y) = N_{B}^{(1)}\left[\cos\left({m^{(1)}_B}y\right) -
\frac{r^a_{B}y}{2}\sin\left({m^{(1)}_B}y\right)\right] , 
\end{equation}
with $N_{W_{3}}^{(1)}$, $N_{B}^{(1)}$ being the normalisation
factors and $r_G^{a},r_G^{b}$,  ($G
\equiv W, B$),  the strengths of the boundary terms
at $y=0$ and $\pi R$  respectively for the gauge bosons.

Further, the five-dimensional gauge couplings $g_5, g_5'$ and the
vacuum expectation value (vev) $v_5$ are related to the usual
couplings $g, g'$ respectively and the vev $v$ defined in four
dimensions through 
\begin{equation}
g_5 = g ~\sqrt{\pi R ~S_W}  \;, \; ~g'_5 = g' ~\sqrt{\pi R ~S_B} \; , \; v_5 =
v/\sqrt{\pi R ~S_H} \;,  
\end{equation}
where
\begin{equation}
S_{W} = \left(1+\frac{R^a_W+R^b_W}{2\pi}\right), \;\; 
S_B = \left(1+\frac{R^a_B+R^b_B}{2\pi}\right),\;\;
S_H = \left(1+\frac{R_h}{\pi}\right).
\end{equation}

A few comments about the mass matrix in eq. (\ref{mmatrix}) may
not be out of place. The matrix is in the $W_3^1 - B^1$
basis\footnote{We have checked that mixing with states of $n \neq
1$ is very small.}.  To estimate the different terms in the
matrix notice that the $S_i$ are ${\cal O}(1)$ as are the overlap
integrals $I_{ij}$. So the contributions to the mass matrix from
the symmetry breaking are ${\cal O}(v^2)$. The order of the
extra-dimensional contributions, ${m_G^{(1)}}^2$, is set by
$(1/R)^2$ and is always much larger by far.  As a consequence to
a good approximation these terms determine the mass eigenvalues
and the mixing  is negligible for $(R_W - R_B) \sim 0.1$ or
larger\footnote{If $R_W = R_B$ then the dominant diagonal terms
are equal and do not contribute to the mixing and simply shift
the masses of the eigenstates. In this case the mixing between
$W_3^1$ and $B^1$ is just as in the Standard Model with $\tan
\theta = g'/g$.}.  So, in our discussion below we take $B^1$ and
$W_3^1$ to be the neutral electroweak gauge eigenstates for the
$n = 1$ KK-level.

\section{Coupling of $B^{1}$ and $W^{1}_3$ with zero-mode fermions}

We have now all the ingredients needed to calculate the coupling
of the states $W_3^{1}$ and $B^{1}$ to two zero-mode fermions
$f^{(0)}$.  Here $f^{(0)}$ could be SM quarks or leptons.  We
separately discuss two cases alluded to earlier. In the first,
the strength of the boundary-localized couplings for fermions
 and Higgs scalars are of the same strength at both fixed
points while for the gauge bosons we allow a variation of the
strengths. In the second case, for the fermion,  the Higgs
scalar, as well as  the gauge bosons we assume that the BLKT are
present at only the $y=0$ fixed point.

\subsection{Symmetric fermion BLKT, general gauge boson BLKT} 

Here we assume for the fermions (Higgs scalars) $r_f^a =
r_f^b = r_f$  ($r_h^a = r_h^b = r_h$) while
for the gauge bosons, $B^1$ and $W_3^1$,  $r^a$ and $r^b$ can be
different in general.  Later, when we examine the prospects at
the LHC, we take $r^a_W \neq r^a_B$ but keep $r^a_W - r^b_W =
r^a_B - r^b_B$.   We choose $R_h \equiv r_h/R = -1.1$. The results are not
very sensitive to the exact value of $R_h$. Our chosen value
ensures that the $H^{1}$ is always heavier than the $B^{1}$ and
$W_3^1$. 

Below we discuss the case for a generic gauge
boson $G^1$, which could be either of $B^1$ or $W_3^1$. The
$y$-dependent wave-functions of our interest here are found to be
\begin{equation}
f_{L}^{0} = g_{R}^{0} = \frac{1}{\sqrt{\pi R(1 + R_f/\pi)}}, 
\end{equation}
and

\begin{equation}
a^{1} = {\cal N}^a _1  
\left[\cos \left( \frac{M_{(1)}y}{R} \right)-\frac{R^{a}M_{(1)}}{2}
\sin \left(\frac{M_{(1)}y}{R}\right)\right],
\end{equation}

with$$
{\cal N}^a _1 = \sqrt{\frac{1}{\pi R}}~\sqrt{\frac{8(4+M_{(1)}^{2}{R^b}^2)}
{2\left(\frac{R^{a}+R^{b}}{\pi}\right)(4+M_{(1)}^{2}R^{a}R^{b})
+(4+M_{(1)}^{2}{R^{a}}^2)(4+M_{(1)}^{2} {R^{b}}^2)}}~~,$$ 

where  we have used as earlier $M_{(1)} \equiv m_{G}^{(1)}  R$, and
the scaled dimensionless variables
\begin{equation}
R_f \equiv r_f/R, \;\;  R^a \equiv r^a_G/R, \;\; {\rm and} \;\;
R^b \equiv r^b_G/R.
\end{equation}
Using the above  we calculate
\begin{eqnarray} 
g_{G^{1}f^{0}f^{0}} &=&  
g_5(G)  ~\int^{\pi R}_{0} 
(1+r_{f}\{\delta(y)+\delta(y-\pi R)\})
f_{L}^{0}f_{L}^{0}a^{1} dy  \nonumber  \\
&=& g_5(G) ~\int^{\pi R}_{0} (1+r_{f}\{\delta(y)+\delta(y-\pi R)\})
g_{R}^{0}g_{R}^{0}a^{1} dy  \nonumber \\
&=& \frac{g(G) \sqrt{S_G} }{\left(1+\frac{R_{f}}{\pi}\right)}
\;{\cal N}^a _1 \;\left[\frac{\sin(\pi M_{(1)})}
{\pi M_{(1)}}\left\{1-\frac{M_{(1)}^{2}R^{a}R_{f}}{4}\right\}\right.
\nonumber \\
&&
\left. +\frac{R^{a}}{2\pi}\left\{\cos(\pi M_{(1)})-1\right\}+
\frac{R_{f}}{2\pi}\left\{\cos( \pi M_{(1)})+1\right\}\right] .
\nonumber \\
&&
\label{coup1}
\end{eqnarray}
The physics consequences of these couplings are discussed in the
next section. We only wish to point out here that it can be readily
seen using eq. (\ref{trans2}) that if $R^a = R^b$, i.e., the BLKTs
are symmetric at $y = 0$ and $y = \pi R$ for {\em both} the
fermion and the gauge boson, then the coupling in eq. (\ref{coup1})
vanishes. This can be traced to  a $[y \longleftrightarrow
(y-\pi R)]$ $Z_2$ symmetry of the theory for this choice which
forbids an $n = 1$ state to couple exlusively to zero modes. 

\begin{figure}[thb] 
\begin{center} 
\includegraphics[width=5cm,height=5.0cm, angle =
270]{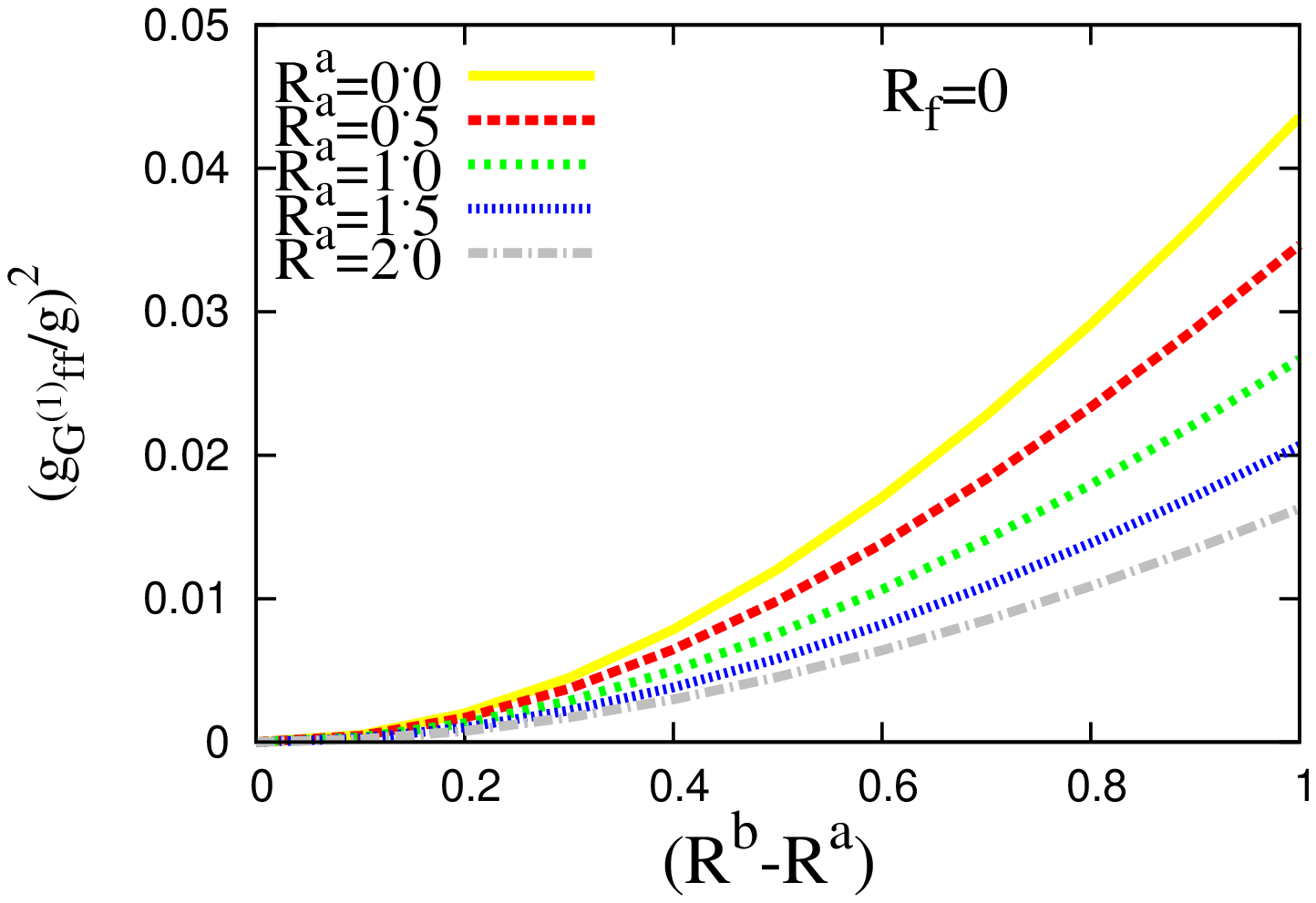}{\hskip 0.5cm}
\includegraphics[width=5cm,height=5.0cm, angle = 270]{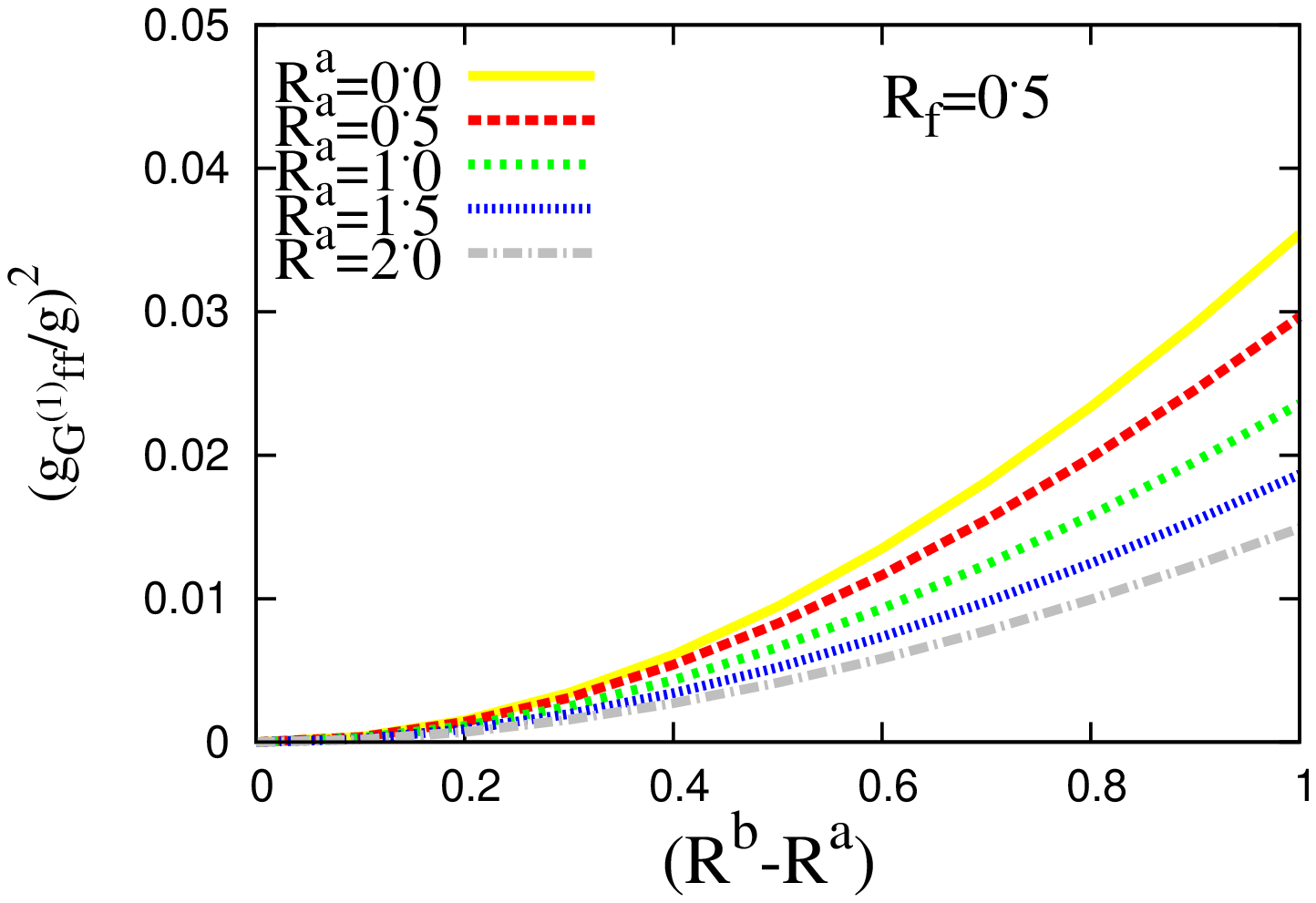}
{\hskip 0.5cm} 
\includegraphics[width=5cm,height=5.0cm, angle =
270]{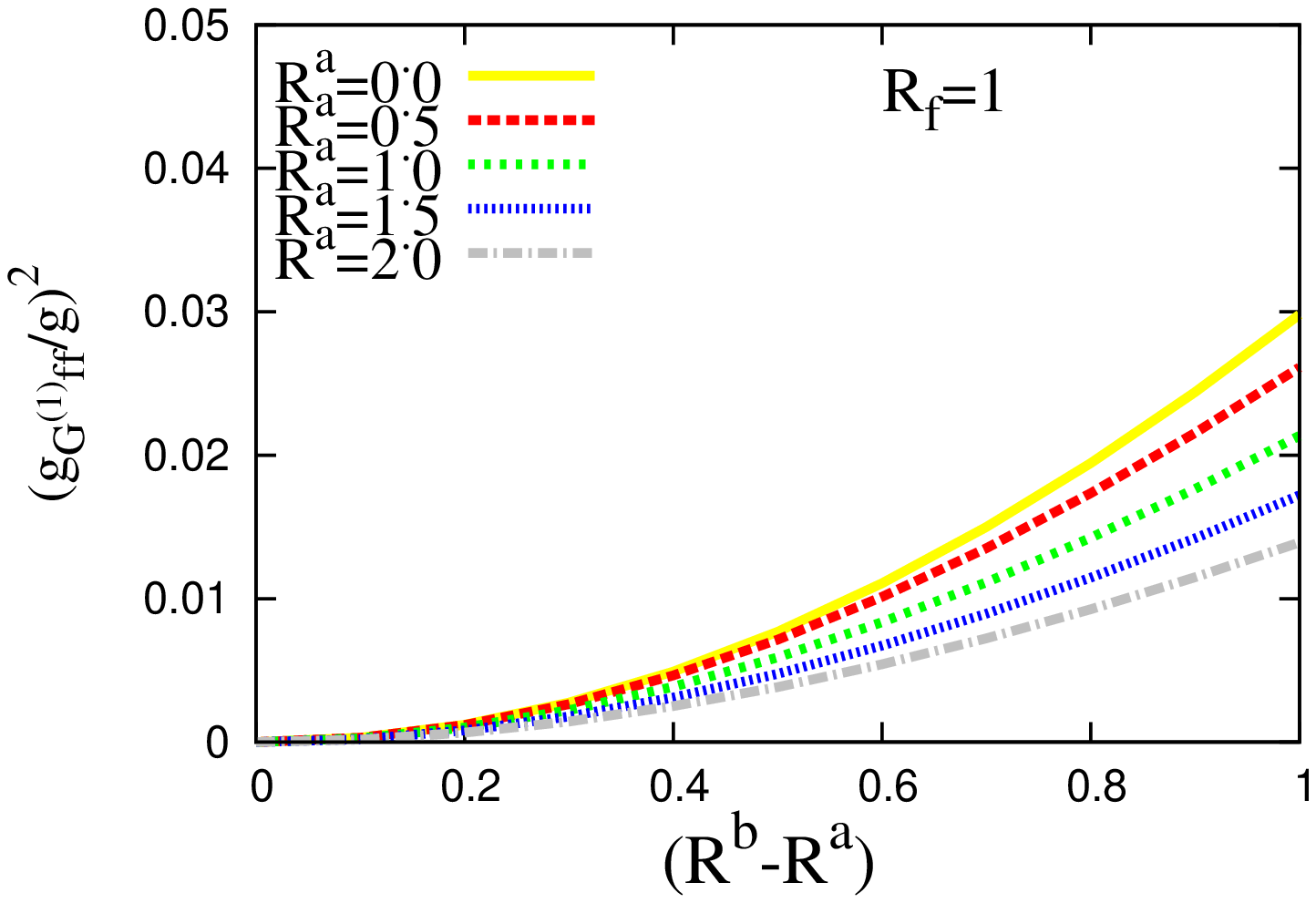}
\caption{ Variation of the square of the KK-parity violating coupling
(between $G^1 \equiv B^1 ~{\rm or} ~W_3^1$ and a pair of zero-mode fermions) 
with $(R^b - R^a)$ for several choices of $R^a$. The panels correspond to
different $R_f$.}
\label{coup_plots} 
\end{center} 
\end{figure}

For illustrative purposes we plot in fig. \ref{coup_plots} the square
of the coupling as a function of $(R^b - R^a)$ for several values of
$R^a$. The  three panels correspond to representative values of
$R_f$. In any of these panels the symmetric BLKT case is obtained when
$\Delta r = 0$.  In this limit the $Z_2$ parity becomes operative and
$G^1$ being odd under this symmetry the coupling vanishes. The choice
$r^a = 0$ implies a set up with the gauge boson BLKT at one fixed
point only while the fermion BLKTs are symmetrically
distributed. 

\subsection{Fermion and gauge boson BLKT at one fixed point only}

\begin{figure}[thb] 
\begin{center} 
\includegraphics[width=5cm,height=4.5cm, angle = 0]{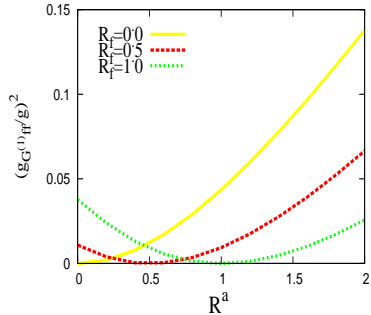} 
\caption{ Variation of the square of the KK-parity violating coupling
(between $G^1 \equiv B^1 ~{\rm or} ~W_3^1$ and a pair of fermions) with  
$R^a$ for different choices of $R_f$ when the fermion and boson
BLKTs are present only at one of the fixed points.}
\label{coup_plotsS} 
\end{center} 
\end{figure}

Now we turn to the case which could be considered the most
asymmetric one, namely, the boundary-localized kinetic terms for
the fermion, the Higgs scalar and the gauge boson are only
at the fixed point $y = 0$ and none at $y = \pi R$. The
$y$-dependent wave-functions in this case are
\begin{equation} 
f_L^{0} = g_R^{0} = \frac{1}{\sqrt{\pi R(1 + R_f/2 \pi)}}, 
\end{equation}
and
\begin{equation}
a^{1} = \sqrt{\frac{1}{\pi R}}~\sqrt\frac{2}{1+\left(\frac{R^a
M_{(1)}}{2}\right)^2+\frac{R^a}{2\pi}}
\left[\cos\left(\frac{M_{(1)}y}{R}\right)-
\frac{R^a M_{(1)}}{2}\sin\left(\frac{M_{(1)} y}{R}\right) 
\right] \;,
\end{equation} 
where $R^a \equiv r^a/R$. With these we find in this case
\begin{eqnarray} 
g_{G^{1}f^{0}f^{0}} &=&  
g_5(G) \int^{\pi R}_{0} \left[1+r_{f} \delta(y)\right]
f_L^{0}f_L^{0}a^{1}\,dy
\nonumber \\ &=&  
g_5(G) \int^{\pi R}_{0} \left[1+r_{f} \delta(y)\right]
g_R^{0}g_R^{0}a^{1}\,dy
\nonumber \\ 
&=&  
\frac{\sqrt{2} ~g(G) \sqrt{S_G}}
{\left(1+\frac{R_{f}}{2\pi}\right) 
\sqrt{1+\left(\frac{R^a M_{(1)}}{2}\right)^2+\frac{R^a}{2\pi}}}  
\left(\frac{R_{f}-R^a}{2\pi}\right) \;.
\label{coup2}
\end{eqnarray} 
In fig. \ref{coup_plotsS} we plot the square of the above
coupling strength as a function of $R^a$ for several choices of
$R_f$. It is seen that the strength of the coupling is roughly of
the same order as in the case discussed earlier. A noteworthy
feature is that the coupling vanishes when $R^a = R_f$.

\section{$B^1$ or $W_3^1$ production and decay} 

We are now in a position to discuss some phenomenological signals
of nmUED. In the following we will restrict our discussion only
to the prospects at the LHC.  Furthermore, our focus will be on
the case where KK-parity is broken as a consequence of having
unequal gauge boson BLKT parameters at the orbifold fixed points.
Though for $B^1$ and $W_3^1$ we choose different BLKT strengths
we keep $(R_G^a - R_G^b)$ to be the same for both so that the
superscript $G$ can be dispensed with in the difference.  We also
explore the situation where the  BLKTs are
present only at one of the two fixed points. Here too we consider
different BLKTs for $B^1$ and $W_3^1$.  The necessary framework for
the analyses has already been set up in the previous sections.

At the LHC we are interested to investigate the resonant
production of the $n=1$ KK-excitations of EW gauge bosons, via
the process $p p$ $(q \bar q)
\rightarrow G^{1}$ followed by $G^{1}
\rightarrow l^{+}l^{-}$ where $G^1$ is either of $B^1$ and $W_3 ^1$ 
and $l^\pm$ could be either $e^\pm$ or $\mu^\pm$. 

From now onwards for the SM particles we will not explicitly
write the KK-number $(n = 0)$ as a superscript.  The particles
with no superscripts are implied to be the SM particles.  
The final state leads to two leptons $(e ~{\rm or} ~\mu)$, with
invariant mass peaked at $m_{G^1}$. It should be noted that both
the production and the decay of $n=1$ KK-excitations of
electroweak gauge bosons are driven by KK-parity violating
couplings which vanish unless the strengths of the BLKT
parameters localized at the two fixed points are different, i.e.,
$(R^b - R^a) \neq 0$.   If in future such a signature is observed
at the LHC, then it would be a good channel for the determination
of such KK-parity violating couplings.

An analytic expression for the production cross section in
proton proton collisions can be written in a compact form :
\begin{equation}
\sigma (p p \rightarrow G^{1} + X) = \frac{4 \pi^2}{3 m_{G^1}^3}\;\sum_i 
\Gamma(G^{1} \rightarrow q_i \bar q_i)\;\int_\tau ^1 \frac{dx}{x}\;
\left[f_{\frac{q_{i}}{p}}(x,m_{G^1}^2) 
f_{\frac{{\bar q_{i}}}{p}}(\tau/x,m_{G^1}^2) + 
q_i \leftrightarrow \bar q_i \right]
\label{x-sections}
\end{equation}

Here, $q_i$ and $\bar{q_i}$ stand for a generic quark and
the corresponding antiquark of the $i$-th flavour respectively.
$\Gamma(G^{1} \rightarrow q_i \bar q_i)$ represents the decay
width of $G^1$ into a quark and antiquark pair of the $i$th
flavour. $\tau \equiv {m_{G^1}^2 / S_{PP}}$, where $\sqrt{S_{PP}}$
is the proton proton centre of momentum energy. The $f$s are  
quark or antiquark distribution functions within a proton.

In case of $B^1$ production, $\Gamma = (g_{G^1 q
\bar{q}}'^2/32 \pi) \left[(Y^{q}_L)^2 + (Y^{q}_R)^2 \right]
m_{B^1}$ (with $Y^q _L$ and $Y^q _R$ being the weak-hypercharges
for the left- and right-chiral quarks), while for $W^1 _3$ one
has $\Gamma = (g_{G^1 q \bar{q}}^2/128 \pi) m_{W^1_3}$.
$g^{(')}_{G^1 q \bar{q}}$ is the KK-parity violating coupling
among SM quarks and $W_3^1$ ($B^1$) as given in eqs.
(\ref{coup1}) and (\ref{coup2}). In the above cross section
expressions $m_{G^1}$ stands for the mass eigenvalue of the gauge
boson $n=1$ excitation resulting from the matrix in eq.
\ref{mmatrix}.

To obtain the numerical values of the cross sections, we
use a parton level Monte Carlo code  with parton distribution functions
as parametrized in CTEQ6L \cite{CTEQ}.  We take the
$pp$ c.m. energy to be 8 TeV.  Renormalisation (for
$\alpha_s$) and factorisation scales (in the parton
distributions) are set at $m_{G^1}$.

To make our estimate of signal cross section more realistic, a simple toy
calorimeter simulation has been implemented with the following
criteria:
\begin{itemize}
\item $p_T ^{\ell} > 20 ~\rm GeV$.
\item The calorimeter coverage  (for leptons) is $\vert
\eta \vert < 3.0$.

\item A cone algorithm with $\Delta R$ = $\sqrt {\Delta\eta^2 +
\Delta\phi^2}= 0.5 $ has been used for isolation of leptons.

\end{itemize}

\begin{figure}[tb]
\begin{center}

\includegraphics[width=5cm,height=7.2cm, angle=270]{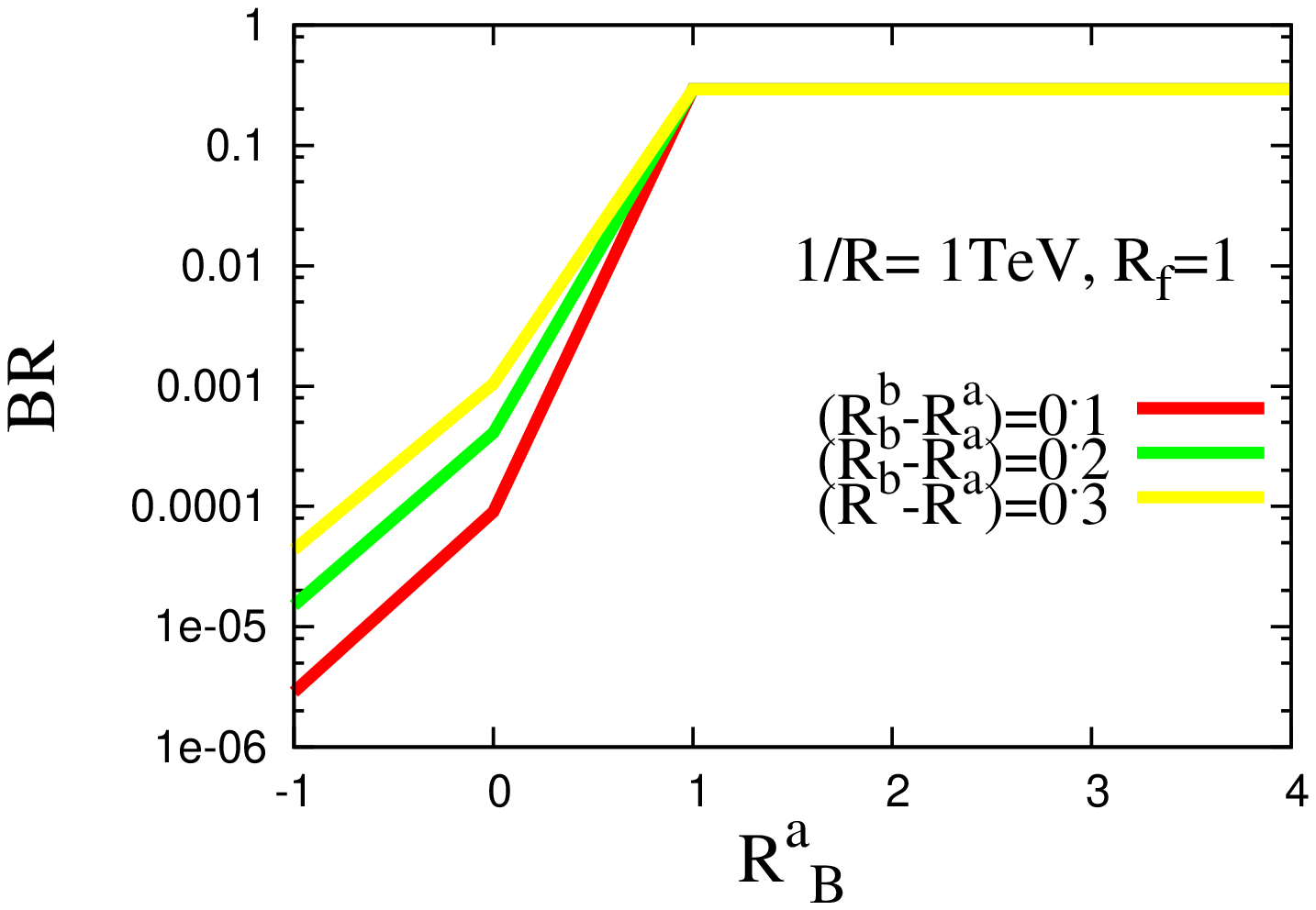}
\includegraphics[width=5cm,height=7.2cm, angle=270]{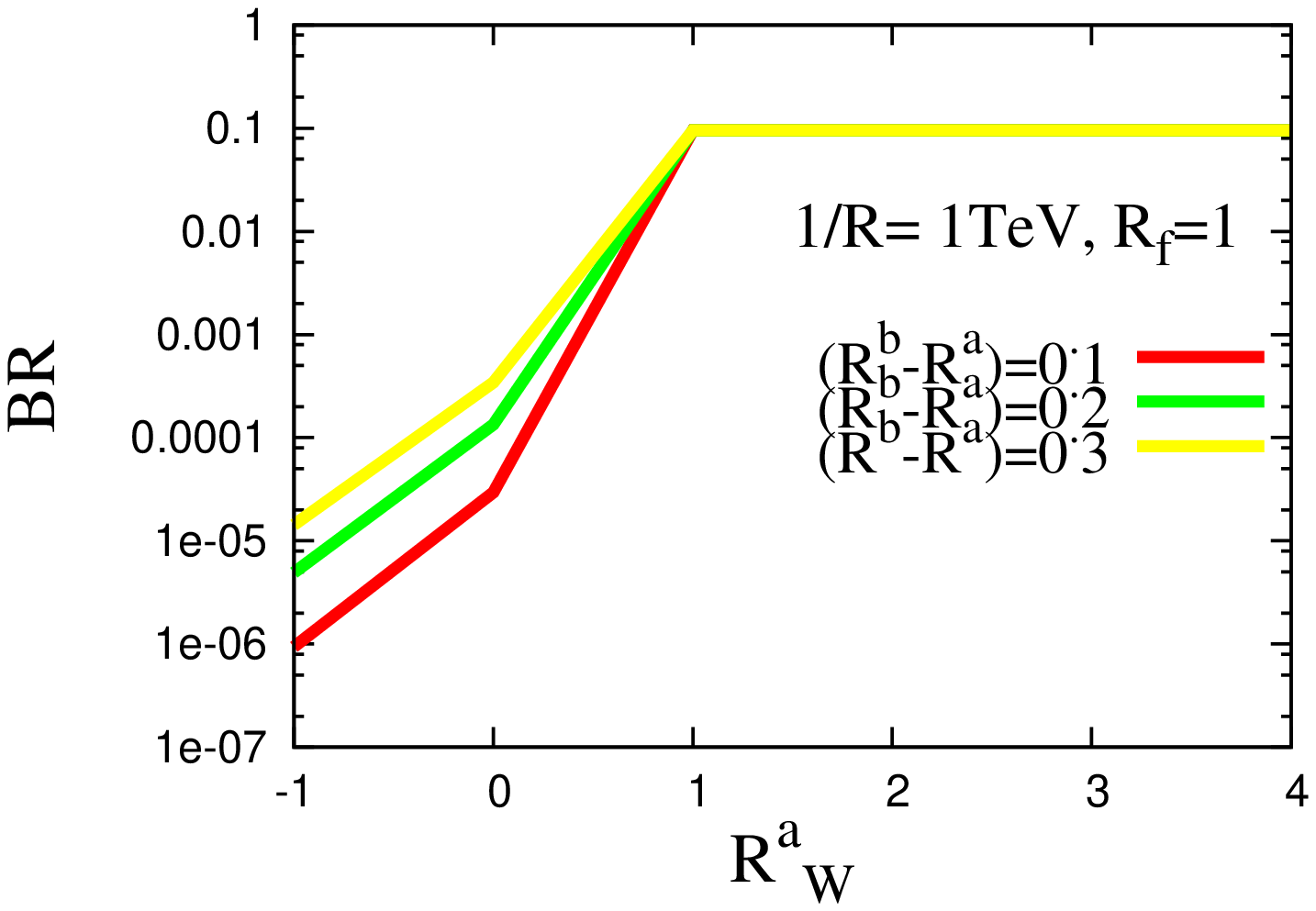}

\caption{ Branching ratios of $B^1$ ($W^1_3$) to $e^+e^-$ or
$\mu^+\mu^-$ as a function of
the BLKT parameters $R^a_B$ ($R^a_W$) are shown in the left
(right panels) for several choices of $(R^b - R^a)$.  $R^{-1} =$ 1
TeV and  $R_f$ = 1 are chosen. Notice that when $R^a_B$ ($R^a_W$)
falls below 1, the $n$ = 1 fermion is lighter and KK-number
conserving decays reduce the BR drastically.}
\label{brD}
\end{center}
\end{figure}

Once produced, $B^1$ ($W_3 ^1$) will decay via similar KK-parity
violating couplings to SM quarks and leptons and, if
kinematically allowed, to $f^1 \bar f$ (or
to $f \bar f^1$) via KK-conserving couplings. We focus on the
KK-parity violating leptonic decays which provide a cleaner
environment at the LHC.  For simplicity we assume an universal
coefficient $r_f$ for the BLKTs involving all SM fermions. If
kinematically allowed (broadly when $R_f > R_G$), the KK-conserving
decay rates can be substantial and in such situations the
branching ratios (BRs) for KK-violating
decay rates are small. This behaviour is
exemplified in fig. \ref{brD} where we plot the BR of the
KK-number violating decay as a function of $R^a_B$ ($R^a_W$) in
the left (right) panel. Note how rapidly (log scale) the BR falls
when the KK-conserving decays open up ($R_f > R^a_{B,W}$).

However, when the KK-conserving decays are kinematically
disallowed, decays to a SM fermion anti-fermion pair are the only
possible modes and hence the branching ratios become independent
of the input BLKT parameters.  Consequently, the decay rates of
$B^1$ to different fermions are proportional to the sum of the
squares of the respective weak hypercharges $\left[(Y^f_L)^2 +
(Y^f_R)^2 \right]$ of the left- and right-chiral species. $W_3
^1$, on the other hand, decays democratically with branching
ratio of $1 \over 21$ to each species of left-handed zero-mode
fermions.  This immediately tells us that the decay branching
ratio of $B^1$ ($W_3 ^1$) to $e^+ e^-$ or $\mu^+ \mu^-$ is
approximately $30\over103$ ($2\over 21$), which is also evident
from the plots in fig. \ref{brD}.

Final states with dileptons arise in
the SM mainly from resonance $Z$-production or Drell-Yan (DY)
production.  The first  of these can be easily vetoed as in this
case the dilepton invariant mass peaks around $m_Z$. We find
that for 10 GeV bins around 700 (800) GeV the DY cross section
is  2.29 (1.56) $fb$. This DY background though non-negligible
is such that for the $W_3^1$ and $B^1$ masses which we consider
$S/\sqrt{B} \geq 5$ can be achieved for 20 $fb^{-1}$ integrated
luminosity.

Before embarking on a detailed exploration it may be useful
to set the perspective by comparing the signal cross section with
the backgrounds noted above. For example, we find that
with $R^{-1}$ = 1 TeV, $R_f = 1$ and $R_W^a = R_B ^a$ = 1.5 which
implies $(R^a_B - R^a_W) = 0$, and $(R^b - R^a)$ = 0.32, one gets
$m_{B^1} (m_{W_3 ^1}) \sim$ 675 (676) GeV and the cross section
for $B_1 (W_3 ^1)$ production is 52.1 (39.1) fb. When the cross
sections are folded with the branching ratios one expects 303
(74) events from $B^1$ ($W^1$) production for 20$fb^{-1}$
integrated luminosity.
Of course, the
$B^1 (W_3 ^1)$ coupling and hence the production rate depend on
the input parameters and it becomes  48.5 (38.5) $fb$ for $R^{-1}$
= 1.0 TeV, $R_f = R_W^a = R_B ^a$ = 1, and $(R^b - R^a)$ = 0.32
when $m_{B^1} (m_{W_3 ^1})
\sim$ 742 (743) GeV leading to 282 (73) events. Therefore we will
not undertake any further effort to estimate the SM background.

\begin{figure}[thb] 
\begin{center} 
\includegraphics[width=5cm,height=7.2cm, angle=270]{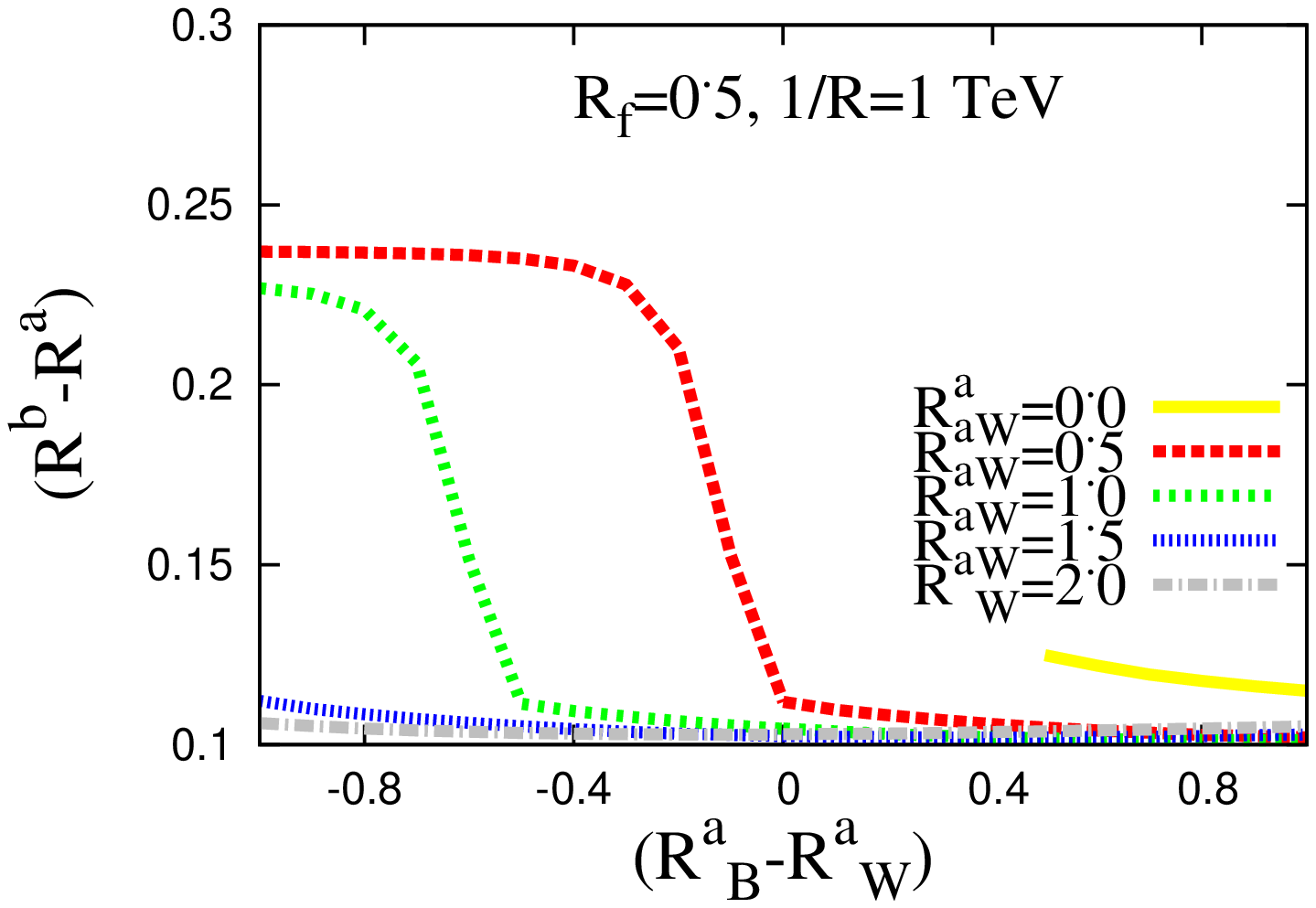}
\includegraphics[width=5cm,height=7.2cm, angle=270]{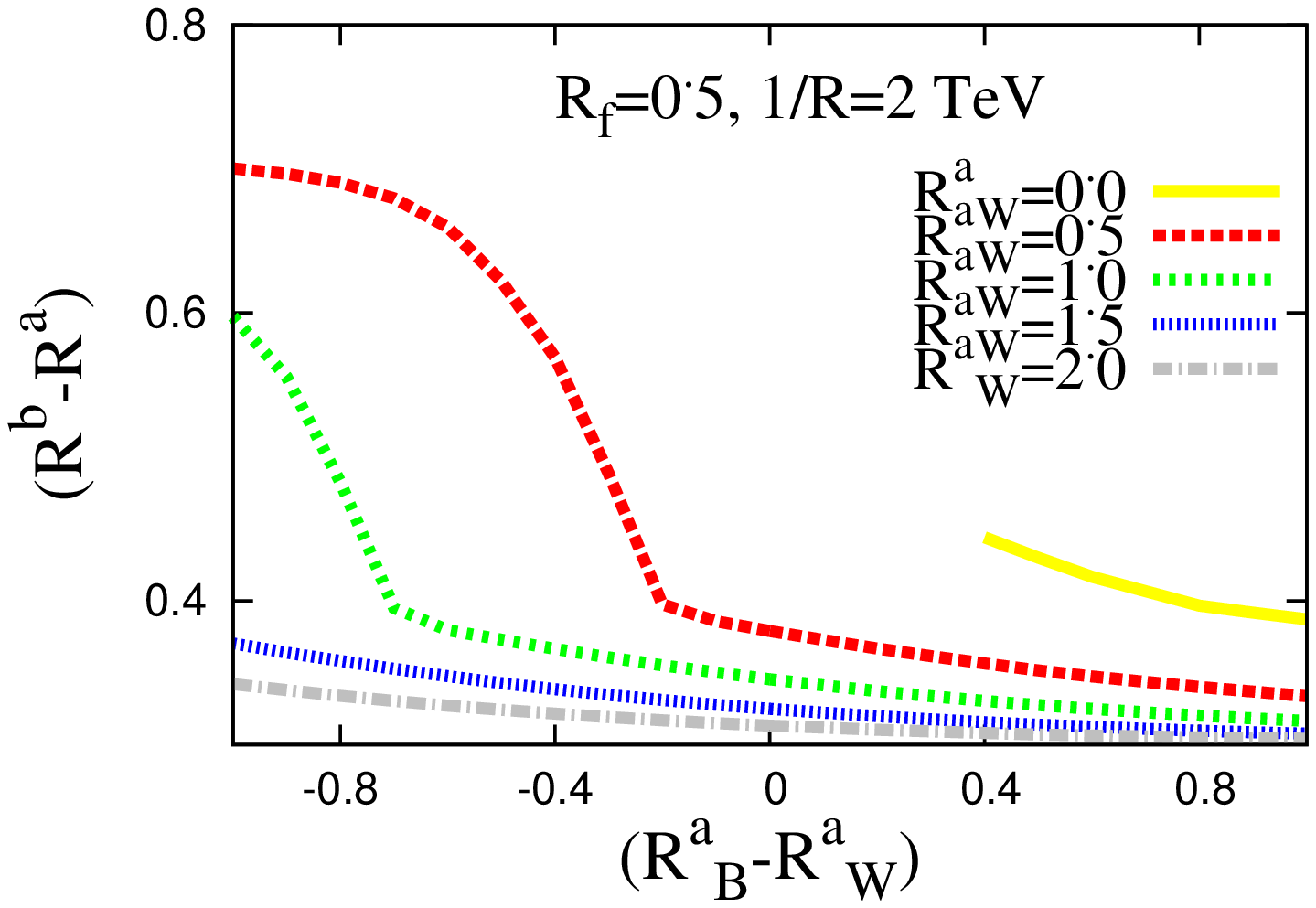}

\includegraphics[width=5cm,height=7.2cm, angle=270]{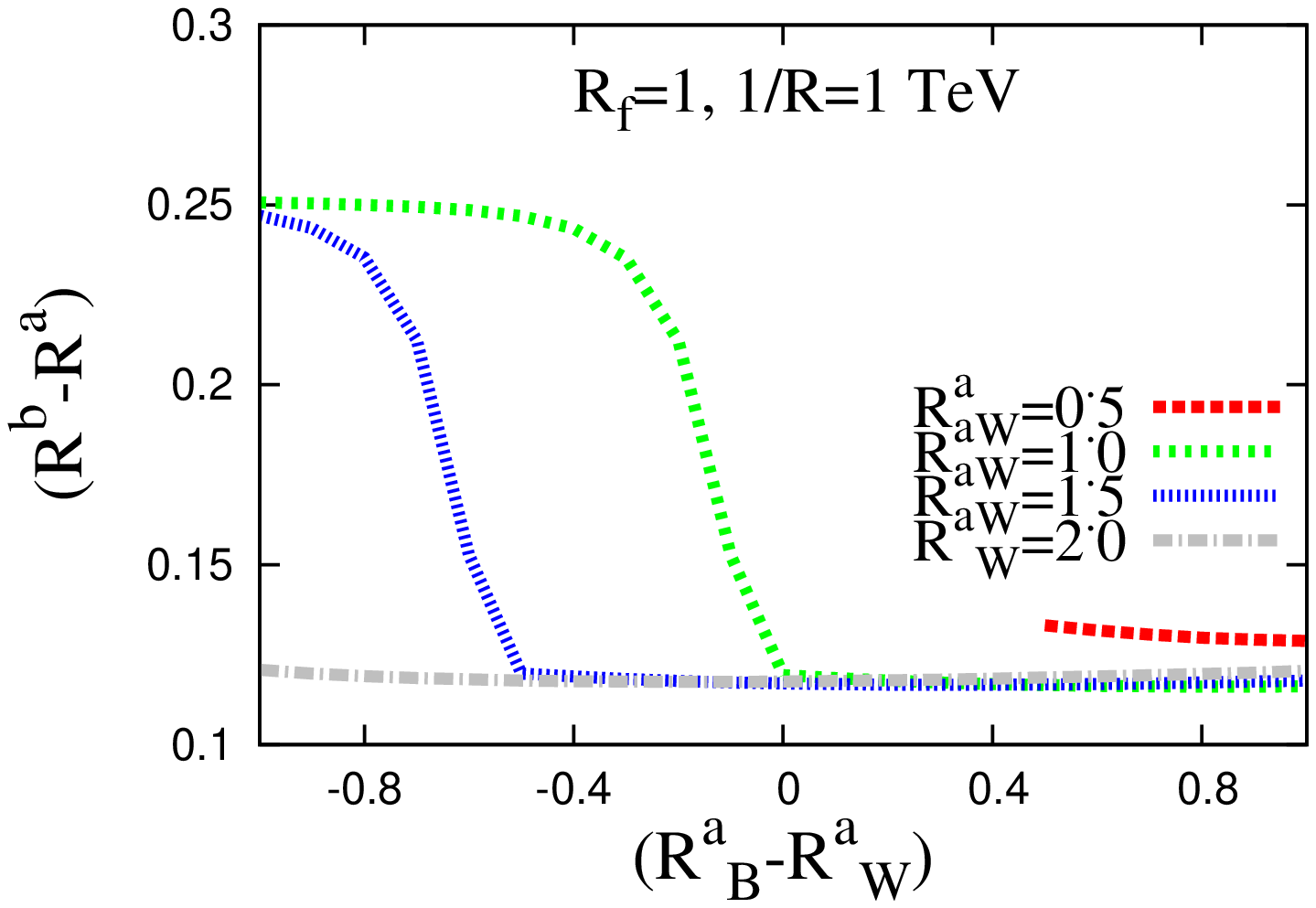}
\includegraphics[width=5cm,height=7.2cm, angle=270]{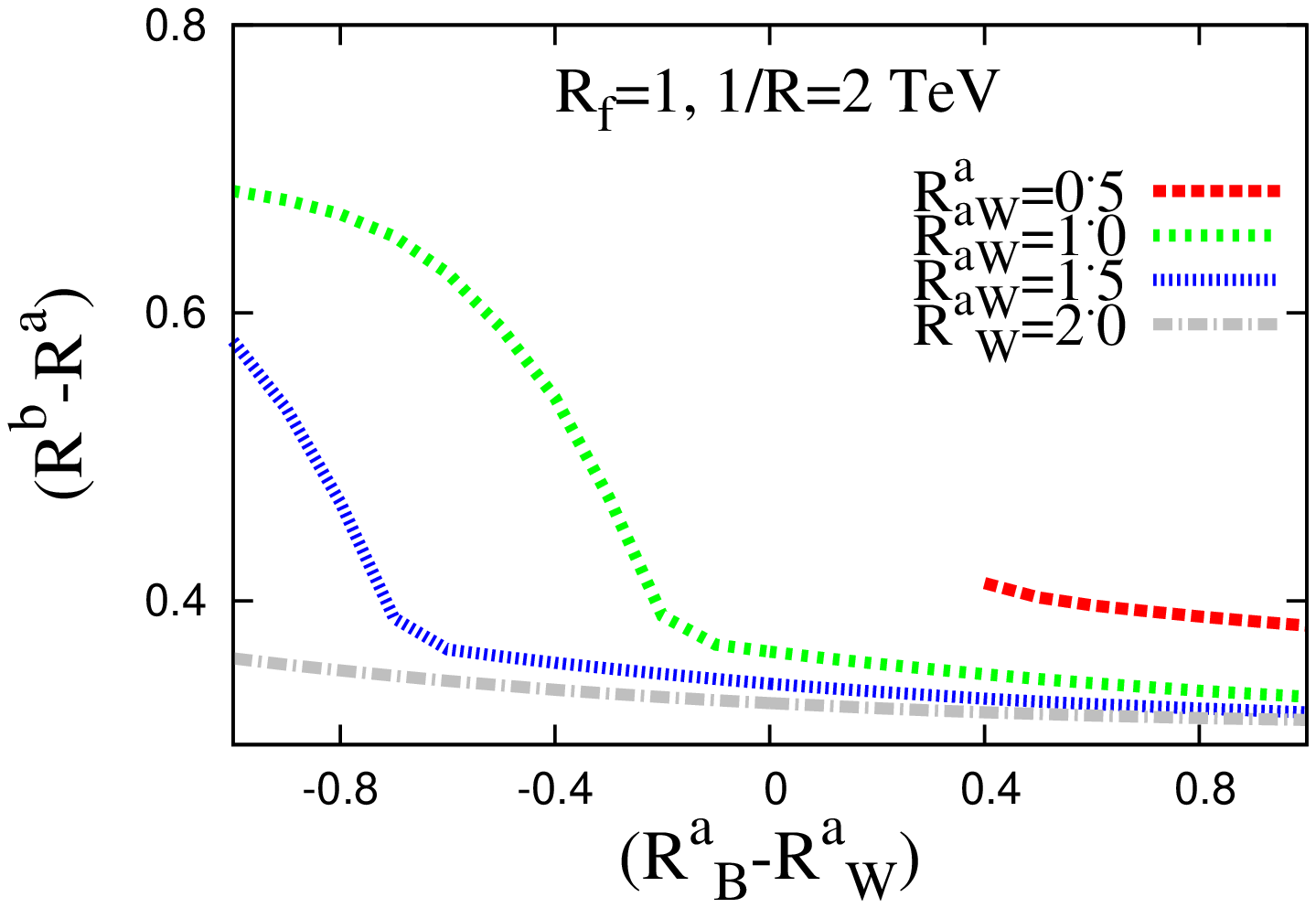}

\caption{Iso-event curves (40 signal events with 20 fb$^{-1}$
data at the LHC running at 8 TeV) for combined $W_3^1$ and $B^1$
signals in the $(R^b - R^a)$ - $(R^a_B - R^a_W)$ plane for
several choices of $R_W^a$ and $R_f$. Each panel corresponds to
specific values of $R_f$ and $R^{-1}$ while each curve in a panel
corresponds to a particular value of $R_W^a$. $(R^b - R^a)$ is
taken to be the same for $W_3$ and $B$. The regions below the
curves correspond to less than 20 events for the chosen $R_f$ and
$R_W^a $. $R^{-1}$ is taken as 1 TeV for the left panels and 2
TeV for the right panels. Note difference in the ordinate scales in the left
and right panels.}
\label{5fb_contours} 
\end{center} 
\end{figure}

We are all-equipped now to present the results.  Rather than
displaying the number of events as a function of the BLKT
parameters,  in Figs. \ref{5fb_contours} and \ref{5fb_contoursS}
we have plotted the iso-event curves\footnote{The requirement
here is that electron {\em plus} muon events together resulting
from $W_3 ^1$ {\em and} $B^1$ production and decay add up to 40.}
(40 events with 20 fb$^{-1}$ luminosity for LHC running at 8
TeV).

\subsection{Symmetric fermion BLKT, general gauge boson BLKT}

Fig. \ref{5fb_contours} is for the case when the fermion BLKTs are
symmetric at the two fixed points while the gauge BLKTs are not.
We show in this case the iso-event curves in the $(R^b - R^a) -
(R^a_B - R^a_W)$ plane\footnote{We have set $(R^b - R^a)$ to be
the same for $B$ and $W_3$ while presenting the results in this
section just to reduce the number of parameters.} for different
choices of $R_W ^a$, $R_f$ and $R^{-1}$. The region above the
curves will be excluded if the signal is not seen at the
projected level. We have checked that the nature of the
results is not sensitive to $R_h$ and we have kept it fixed at
-1.1 throughout. In this manner we evade the possibility of the
Higgs excitation being the LKP.

At the very outset, we point out that for similar values of input
parameters, the 2-lepton rate from $W_3 ^1$ production and decay
is smaller than that from $B^1$ approximately by a factor of 4.

We now illustrate a feature of the curves in fig.
\ref{5fb_contours} taking as reference the top
left panel ($R_f = 0.5$, $R^{-1} = 1$ TeV). The curve for $R^a_W
= 0$ has a different nature from the other curves. For large
values of  $(R^a_B - R^a_W)$, where it exists at all, the curve
is away from the other ones which essentially overlap in this
region and also, unlike the other curves, it  terminates sharply
at a value of $(R^a_B - R^a_W) \simeq 0.5$. This behaviour can be
understood with reference to the $n$ =1 KK state masses as shown
in fig. \ref{KKmass}.  For $(R^a_B - R^a_W) \leq 0.5 $, beyond
where the curve terminates, the fermion $n = 1$ state ($R_f =
0.5$, $R^a_W = 0$) becomes the LKP and so this region is
excluded. At larger values the $B^1$ state is the LKP and is
admissible.  The other point is that with $R^a_W = 0$ and for the
small range of $(R^b - R^a)$ allowed, the $n =1$ fermion states
are lighter than $W^1_3$ which is unlike the rest of the curves
of the panel. So, for these latter curves when $R^a_B$ becomes
less than 0.5 the $B^1$ ($W_3^1$) dominantly decays through
KK-conserving (-violating) channels, resulting in the jump in the
curves.

Increasing $(R^b - R^a)$ pushes the $B^1$ and $W_3 ^1$ production
cross sections monotonically upwards, hence allowing us to probe
higher and higher  $R^{-1}$.  Dependence on $R^{-1}$ creeps into
the cross section via the mass of the KK-gauge bosons. As noted,
this extra-dimensional contribution dominates over the part due
to gauge symmetry breaking. A higher $R^{-1}$ implies an enhanced
$m_{G ^1}$ which in turn decreases the cross section.  This
explains the larger $(R^b - R^a)$ needed to have the same signal
rate when we go from the left panels ($R^{-1}$ = 1 TeV) to the
right panels ($R^{-1}$ = 2 TeV) of fig.
\ref{5fb_contours}.  Thus, $(R^b - R^a)$ and
$R^{-1}$ affect the $G^1$ production cross section in opposite
directions.  Increasing the fermion BLKT, $R_f$, would result in
diminshing the coupling between the SM fermions with the
KK-EW-gauge boson (see fig. \ref{coup_plots}). Thus higher $(R^b
- R^a)$ is necessary to compensate the loss in production rate.
This feature is clearly reflected in the difference between the
plots in the top and bottom rows of fig.
\ref{5fb_contours}.

Let us finally focus on the dependence on
$R_W^a$.  As already pointed out (see fig. \ref{KKmass}), larger
values of gauge BLKT results in lower KK-gauge boson masses, thus
pushing the production rate upward.  This explains the lower
values of $(R^b - R^a)$ -- hence smaller value of KK-gauge-SM
fermion couplings -- required to maintain the same production
rate in fig. \ref{5fb_contours} when one compares the different curves in any
one  panel.

Data have been collected at LHC  at 8 TeV
proton-proton c.m. energy.  Observation or otherwise of
the proposed  dilepton signal would 
help in exploring or excluding the parameter space of such
non-minimal UED models.  The projected exclusion limits can be
directly read off from the plots in fig. \ref{5fb_contours}.
Any point above a particular iso-event curve can
be excluded from the non-observation of such events.  In the two
fixed point set up, the cross section has a mild dependence on
$R_f$ and $(R^a_B - R^a_W)$.  The $r_f$ dependence of the cross
section comes only through the coupling.  A careful examination
of the coupling in eqs.  (\ref{coup1}) and (\ref{coup2}) shows
that it tends to a constant as $R_f
\rightarrow \infty$.

\subsection{Fermion and gauge boson BLKT at one fixed point only}

\begin{figure}[tb]
\begin{center}

\includegraphics[width=5cm,height=7.2cm, angle=270]{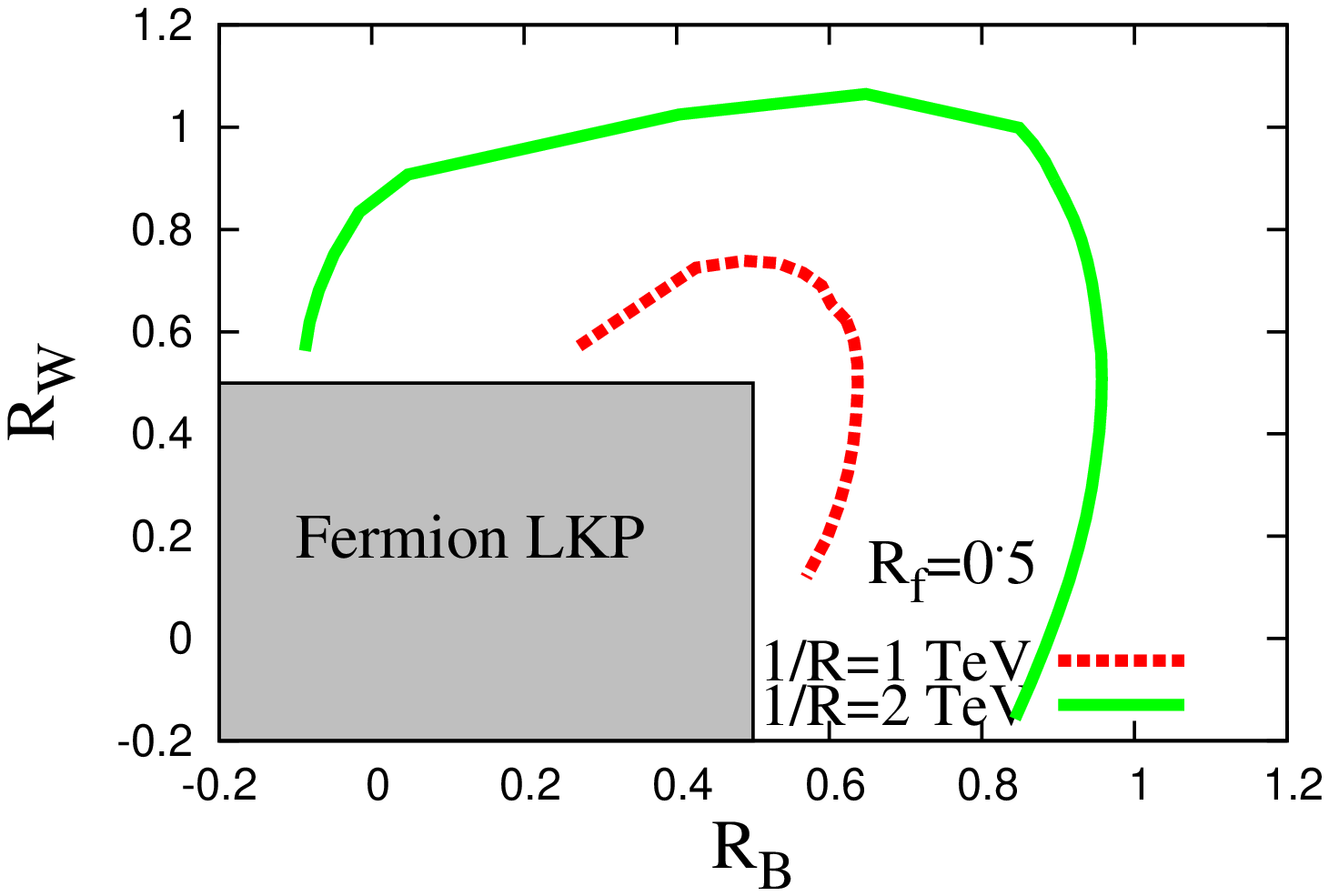}
\includegraphics[width=5cm,height=7.2cm, angle=270]{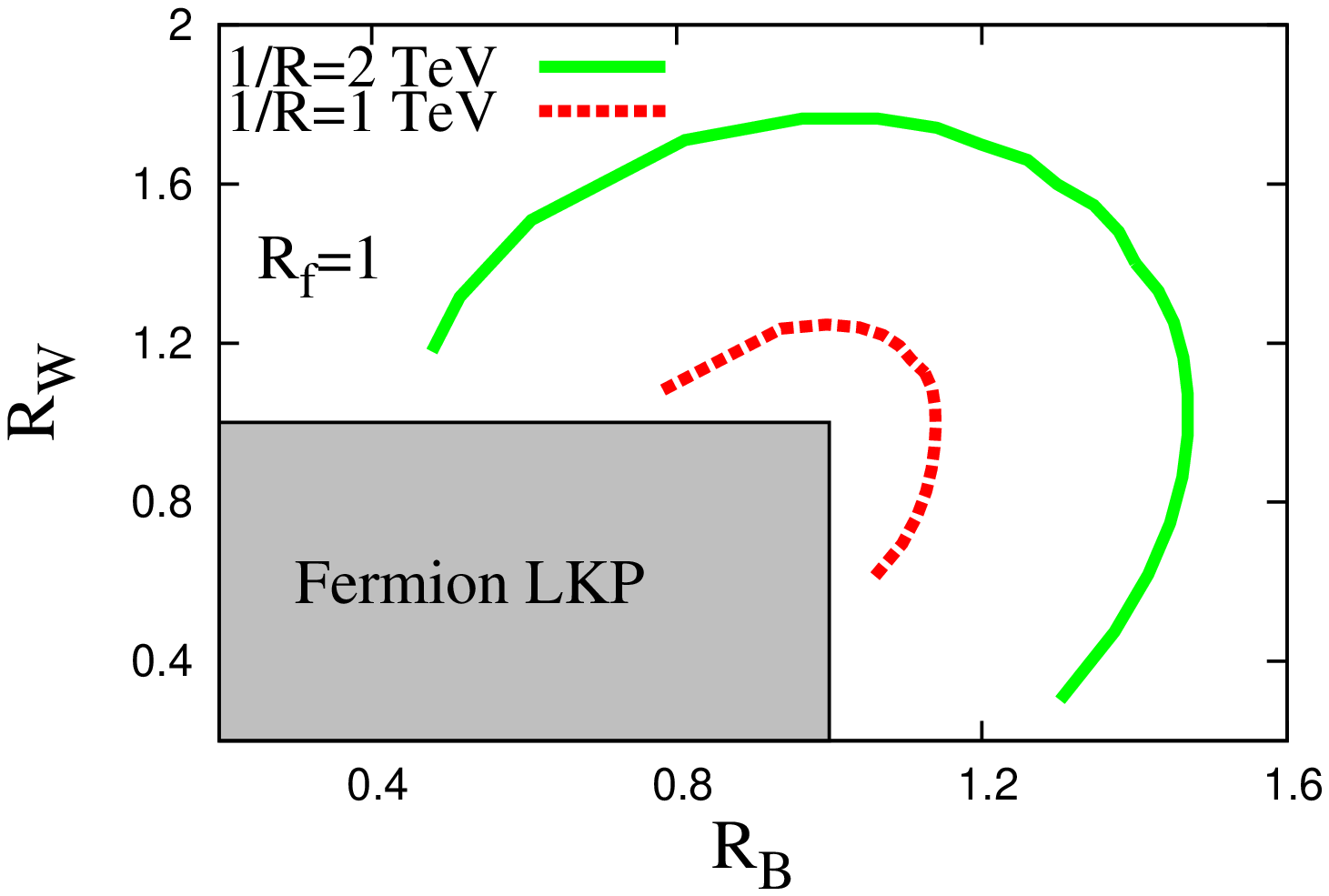}

\caption{ Iso-event curves (40 signal events with 20 fb$^{-1}$
data at the LHC running at 8 TeV) for combined $W_3^1$ and $B^1$
signals in the $R_W - R_B $ plane when the BLKTs are present at
only one of the two fixed points. The region enclosed by the
solid green (dashed red) curve and the fermion LKP box yields
less than 20 events for $R^{-1} =$ 2 TeV (1 TeV). The left (right)
panel corresponds to  $R_f$ = 0.5 (1.0). Note difference in scales
in the two panels.}
\label{5fb_contoursS}
\end{center}
\end{figure}

Now let us turn to the case of BLKTs at only one fixed point
(fig. \ref{5fb_contoursS}).  The relevant KK-parity violating
couplings vanish when\footnote{Since for this option the BLKTs
are present at only one fixed point we denote them by $R_f,
~R_W$, and $R_B$.} $R_f = R_G$ where $G \equiv W$ or $B$ as noted
from  eq. (\ref{coup2}). This is seen in fig.  \ref{coup_plotsS}
for different choices of $R_f$. So, the production mode we
consider here is not available. Further, in this situation, the $n$
= 1 gauge boson and  fermion states are mass degenerate and
KK-number conserving decay modes are also not allowed.  In the
neighbourhood of 
this point there is an important asymmetry, however, between
whether $R_{B,W}$ is more than $R_f$ or less. In the
former case, the gauge boson $n$ =1 state is lighter than the
corresponding fermion state and KK-number conserving decays are
not kinematically possible. As shown in fig.
\ref{brD} the KK-number violating decay branching ratio is high.
In the latter case, the  KK-number conserving decay is allowed
for the $n$ = 1 gauge boson and so the desired branching ratio to
zero-mode fermions is small. Of course, if both $R_{B}$ and
$R_{W}$ are smaller than $R_f$ then the fermion state is the LKP.

The evidence of the
above observations can be readily found in the two panels of
fig. \ref{5fb_contoursS}. For any of the iso-event contours, the
largest value of $R_B$ will be for $R_W$ = $R_f$ and {\em
vice-versa}. This is because when $R_W$ = $R_f$ there is no
contribution to the signal from the $W^1_3$ as discussed above.
As $R_W$ shifts from this value the contribution from $W^1_3$
reduces the needed $R_B$. But this is not symmetric because of
the difference in the branching ratios of $W^1_3$ on the two
sides of $R_f$ as  is seen from  Fig. \ref{brD}. 

We have presented the iso-event
curves (again 40 events with 20 fb$^{-1}$ luminosity) for this
scenario in the $R_W^a - R_B^a$ plane.  We also show the excluded
region where the $n$ = 1 fermion is the LKP. In accordance with
the preceding discussion, it is readily seen that for the point
$R_W = R_B = R_f$ the KK-number violating coupling is zero and
also the $n=$1 states are all degenerate. So, neither KK-number
violating nor conserving decays are permitted.

\section{Conclusions} 
To summarize, we have considered the effects of boundary
localized kinetic terms  in a situation where all the SM
fields can propagate in a spacetime with four spatial and one
timelike dimensions. The extra spatial dimension $y$ is compact
and can be considered as a circle of radius $R$ and also has a $y
\leftrightarrow -y$ symmetry. Consequently there are two fixed
points at $y = 0$ and $y = \pi R$. At these boundary points one
can include terms consistent with 4-dimensional Lorentz symmetry.
These are either kinetic terms or mass terms. We concentrate on
the former.

In the minimal Universal Extra Dimensional model, radiative
corrections play a  crucial role in removing the near-degeneracy
of the masses of the KK-modes of  all SM particles of the same
level, $n$.  UED, being an effective theory, is defined with a
cut-off, $\Lambda$.  In mUED the boundary terms are  chosen in a
manner such that  at $\Lambda$ the contribution due to radiative
corrections vanishes. In this process, instead of calculating the
radiative correction in a 5d set up one can also parametrise
these effects by incorporating a set of BLKTs.  

There are two possibilities of choosing the BLKTs with rather
distinct physics consequences. In the first, the BLKTs are of
equal strength at both the boundary points $(y = 0, \pi R$).
Here, a $Z_2$ symmetry $y \longleftrightarrow (y - \pi R)$ remains.
One then ends up with a theory where the spectrum of KK-particles
and the couplings can be drastically
different from mUED. The lightest among the $n = 1$ KK particles can be a
dark matter candidate.  The other alternative is to allow the
BLKTs at $y=0$ and $y = \pi R$ to be of unequal strengths.
This  will lead to a breakdown of KK-parity and will allow, for
example, the decay we have examined, $B^1 (W_3^1) \rightarrow
e^+e^-, \mu^+\mu^-$, and production of the $B^1 (W_3^1)$ singly. 

In this article, we have considered the possible BLKTs for an
interacting theory of {\em only} fermions and the neutral
electroweak gauge bosons.
In one alternative, the strengths of fermion BLKTs at the two
fixed points have been assumed to be equal $\equiv r_f$. For the
gauge boson boundary kinetic terms we have considered the general case of
unequal BLKTs $(r_G^a \neq r_G^b)$. Equality of the latter strengths
would restore a $Z_2$-parity. As an alternate possibility we have
considered the situation where the fermion and gauge boson BLKTs are
present {\em only} at the $y = 0$ fixed point. Presence of the
boundary terms modify the field equations in the $y$-direction.
Consistency conditions of the solutions of the above equations
lead to the  masses of KK-excitations of fermions and the photon.

For the purpose of illustration, we have calculated the coupling
of $W_3^1$ and $B^1$, the $n = 1$ KK-excitations of the neutral
electroweak gauge bosons, to a
pair of zero-mode fermions (i.e.,  SM fermions) as a function of
$r_f, r^a, r^b ~{\rm and} ~R^{-1}$.  In general, we have
presented the couplings as a function of the scaled variable
$(R^b - R^a)$  for several choices of the other parameters. This
coupling is a hallmark of KK-parity violation and vanishes in the
$(R^b - R^a) = 0$ limit.  A similar KK-parity violating
coupling, which arises  when the BLKTs are present only at $y
=0$,  has also been evaluated. Finally, the production and decay
of $W_3^1$ and $B^1$ at the LHC, via the above KK-parity violating
coupling, have been considered. We have investigated the
viability of the {\em dilepton} signature at the LHC
running at 8 TeV $pp$ center of mass energy.  It is
revealed that non-observation of such a high mass {\em dilepton}
signal  with 20 fb$^{-1}$ accumulated luminosity in the 8 TeV run
of LHC will disfavour a large part of the parameter space
(spanned by $r_f, r^a, r^b ~{\rm and} ~R^{-1}$).

This particular new physics signal can also arise if there are
extra $Z$-like bosons in the context of extensions of the SM,
e.g., the Left-Right  symmetric models or models with an extra
$U(1)$ symmetry. We have not attempted to compare the signals of the
model under consideration with those in these other scenarios.

Let us briefly comment here on the numerical
values of some of the input parameters. In ref.
\cite{delAguila_STU} constraints on the BLKT parameters have
been derived from the consideration of electroweak precision
variables $S$, $T$, and $U$.  In this work  the authors have
considered a particular choice of gauge BLKT parameters, namely,
they have set $R_W^a = R_B^a$. As pointed out in an
earlier section, this choice of BLKT would lead to the same
mixing at all levels $n$ as for the zero modes, i.e.,  
the weak mixing angle (28$^0$) in each KK-level.
On the other hand, in general, $\theta _W ^n$ is negligibly small
for unequal  $U(1)$ and $SU(2)$ BLKTs.  It is well known that the
value of $\theta_W$ plays a crucial role in the exercise done
with precision observables and so the constraints derived  in
\cite{delAguila_STU} are not directly applicable here.

Our analysis is simple-minded in the sense that we have
not considered the effects of initial and final state radiations,
showering and detector effects while estimating the signal
cross section.  Our demarcations of the excluded regions should
therefore be considered only as indicative.

{\bf Acknowledgements} AD acknowledges partial support from the
DRS project sanctioned to the Department of Physics, University
of Calcutta by the University Grants Commission. UKD is supported
by  funding from the Department of Atomic Energy, Government of
India for the Regional Centre for Accelerator-based Particle
Physics, Harish-Chandra Research Institute (HRI).   AS is the
recipient of a Junior Research Fellowship from the University
Grants Commission. AR is thankful to the Department of Science
and Technology for a J.C.  Bose Fellowship.

\end{document}